\documentclass[aps,prb,groupedaddress]{revtex4}

\usepackage{graphicx}
\usepackage{dcolumn}
\usepackage{epsfig}
\usepackage{bm}
\usepackage{amsfonts}
\usepackage{latexsym}

\newcommand{\be}{\begin{equation}}
\newcommand{\ee}{\end{equation}}
\newcommand{\bea}{\begin{eqnarray}}
\newcommand{\eea}{\end{eqnarray}}

\newcommand{\Tr}{{\rm Tr}}

\newcommand{\figwidths}{0.35\columnwidth}
\newcommand{\figwidth}{0.45\columnwidth}

\begin{document}


\title{Finite temperature strong-coupling expansions for the Kondo lattice model}



\author{J. Oitmaa}
\email[]{j.oitmaa@unsw.edu.au}
\affiliation{School of Physics,
The University of New South Wales,
Sydney, NSW 2052, Australia.}

\author{Weihong Zheng}
\email[]{w.zheng@unsw.edu.au}
\homepage[]{http://www.phys.unsw.edu.au/~zwh}
\affiliation{School of Physics,
The University of New South Wales,
Sydney, NSW 2052, Australia.}

%

\date{\today}

\begin{abstract}
Strong-coupling expansions, to order $(t/J)^8$, are derived for the
Kondo lattice model of strongly correlated electrons, in 1-,
2- and 3- dimensions at arbitrary temperature. Results are
presented for the specific heat, and spin and charge susceptibilities.
\end{abstract}


\maketitle

\section{\label{sec:intro}INTRODUCTION}

This paper, the second of a sequence, studies the thermodynamic
properties of the Kondo lattice model, described by the Hamiltonian
\be
H = -t \sum_{\langle ij\rangle \sigma } ( c_{i\sigma}^{\dag} c_{j\sigma} + h.c.) 
+ J \sum_i {\bf S}_i \cdot {\bf s}_i - \mu \sum_{i\sigma} n_{i\sigma}
\ee
The first term describes a single band
of conduction electrons, the ``Kondo coupling" term
represents an exchange interaction between  conduction
electrons and a set of localized $S=\frac{1}{2}$ spins, and
the final term allows for variable conduction electron
density via a chemical potential.

The Kondo lattice model combines two competing physical effects.
In the strong coupling (large $\vert J\vert$) limit, the
conduction electrons will form local singlets ($J>0$) or triplets
($J<0$) with the localized spin at each site. In either case
there will be a gap to spin excitations and spin correlations will be 
short ranged. On the other hand, at weak coupling, the
conduction electrons can induce the usual RKKY interaction
between localized spins, leading to magnetic order.

Despite the apparent simplicity of the model, no exact results
are known, either at $T=0$ or at finite temperatures,
in any spatial dimension. In the preceding paper\cite{zwh}
we studied the  ground state properties of the model at $T=0$, using linked-cluster
series expansions. We refer to that paper for a discussion of other work, which
has been, almost exclusively, restricted also to $T=0$.
In the present paper we focus on finite temperature thermodynamic
properties. We know of only a few previous studies of this kind.
R\"oder {\it et al.}\cite{rod97} considered the ferromagnetic model in the
limit $J\to -\infty$, on the simple cubic lattice, via a high
temperature expansion. Here we treat the general $J$ case and focus on
the antiferromagnetic model.
Shibata {\it et al.}\cite{shi} have studied the 1D antiferromagnetic model via 
a finite-temperature DMRG approach.
 Haule {\it et al.}\cite{hau00} have
treated the 2D case, primarily via a numerical finite 
temperature Lanczos method. 
We compare our results with this work wherever possible.
Haule {\it et al.} have also considered the
atomic limit ($t=0$), and the order $t^2$ correction terms.
Our work was largely motivated by this paper.

Our approach, which will be described in the following Section,
treats the single site terms exactly and treats the hopping
term perturbatively. It is, thus, an expansion about the
``atomic limit". We summarize here, for  completeness and
for later  reference, the exact results in this limit.

For variable conduction electron density there are  8 states per 
site: 2 states with no conduction electrons and localized
spin up or down, 2 states with two conduction electrons of
opposite spin, and 4 states (a singlet and 3 triplets)
with one conduction electron coupled to the localized spin.
For a lattice of $N$ sites the grand partition function is
\be
{\cal Z}_0 \equiv \Tr \{ e^{-\beta H} \} = z_0^N \label{eq1}
\ee
with
\be
z_0 = 2 + (e^{3 K} + 3 e^{-K}) \zeta + 2 \zeta^2 
\ee
and $K\equiv \beta J/4$, $\zeta = e^{\beta \mu}$, $\beta = 1/{k_B T}$.

The internal energy per site is given by
\bea
u (\zeta, T) &=& - {\partial \over \partial \beta}{\Big \vert}_{\zeta} \ln z_0\nonumber \\
  &=& {-\frac{3}{4} J (e^{3 K} - e^{-K}) \zeta \over
  2 + (e^{3K} + 3 e^{-K}) \zeta + 2 \zeta^2 }  \label{eq3}
\eea
The fugacity can, as usual, be eliminated in favour of the
electron density $n$ by using the relation
\bea
n (\zeta, T) &=& \zeta {\partial \over \partial \zeta}{\Big \vert}_T \ln z_0 \nonumber \\
  &=& {q \zeta + 2 \zeta^2 \over 1 + q \zeta + \zeta^2 } \label{eq4}
\eea
where we have introduced $q \equiv \frac{1}{2} (e^{3 K} +3 e^{-K})$.
Solving this gives
\be
\zeta = {- q (1-n) + \sqrt{ q^2 (1-n)^2 + 4 n (2-n) } \over 2 (2-n) } \label{eq5}
\ee
The specific heat is then obtained from (\ref{eq3}) and (\ref{eq5}) via
the usual relation $C_v={d u\over d T}$.
 
Also of interest are the compressibility or ``charge susceptibility"
\be
\chi_c = {\partial n \over \partial \mu}
\ee
which can be expressed as
\be
\beta^{-1} \chi_c = {q (1-n) \zeta + 2 (2-n) \zeta^2 \over 1 + q \zeta + \zeta^2}
\ee
and the magnetic susceptibility, which is given by
\be
4 \beta^{-1} \chi_s = {1 + 4 e^{-K} \zeta + \zeta^2 \over 1 + q \zeta + \zeta^2} \label{eq7}
\ee
Simpler analytic results can be obtained in various limits:
at high or low temperature, and at or near half-filling. Some of these are given in
Ref. \onlinecite{hau00}, although they contain minor errors in a few
cases.

Of course in the atomic limit all of these quantities are smooth functions of 
temperature and electron density.

\section{\label{sec2}Thermodynamic Perturbation Theory}
Our goal is to obtain an expression for the thermodynamic potential,
and hence other quantities, in powers of $t/J$. We work in the grand
canonical ensemble and write the Hamiltonian as
\be
H = H_0 + V
\ee
with 
\bea
H_0 &=& J \sum_i {\bf S}_i \cdot {\bf s}_i - \mu \sum_{i\sigma} n_{i\sigma}  \\
V &=& - t \sum_{\langle ij\rangle \sigma} (c_{i\sigma}^{\dag} c_{j\sigma} + {\rm h.c.})
\eea

The grand partition function can then be expanded in the usual way as
\bea
{\cal Z}_g  &=& \Tr \{ e^{-\beta (H_0 + V) } \} \nonumber \\
&=& {\cal Z}_0 {\Big \{} 1 + \sum_{r=1}^{\infty} (-1)^r \int_0^{\beta} d \tau_1
\cdots \int_0^{\tau_{r-1}} d \tau_r \langle \tilde{V} (\tau_1) \cdots \tilde{V} (\tau_r)\rangle {\Big \}}
\eea
with
\be
\tilde{V} (\tau) = e^{\tau H_0} V e^{-\tau H_0}
\ee
and 
\be
\langle A \rangle = {1\over {\cal Z}_0} \Tr \{ e^{-\beta H_0} A\}
\ee
where ${\cal Z}_0$ is the atomic limit partition function (\ref{eq1}).
The free energy (grand potential) is then given by
\be
-\beta F = N\ln z_0 + \sum_{r=1}^{\infty} (-1)^r T_r
\ee
where
\be
T_r = \int_0^{\beta} d \tau_1 \cdots \int_0^{\tau_{r-1}} d \tau_r
 \langle \tilde{V} (\tau_1) \cdots \tilde{V} (\tau_r)\rangle_N \label{eq13}
\ee
and the subscript $N$ signifies that only the part proportional to $N$ is
to be included.

This approach is, of course, well known and has been used in the past
for both pure spin models\cite{els98} and for the Hubbard model\cite{hen92}.
Any contribution to $T_r$ in (\ref{eq13}) comes from a particular cluster of
sites and bonds, a ``graph". It is possible to restrict the class of graphs to 
connected ones only, as done in Ref. \onlinecite{els98}. In our work
we included also disconnected graphs, of more than one component, as
they are rather easy to deal with directly. Since each bond contributes
a $\tilde{V}$ operator and hence a factor $t$, it is obvious that to carry
the expansion to order $t^r$ all topologically distinct graphs with
up to $r$ bonds need to be considered. There are a total of 115 graphs through
8th order, which is as far as we have been able to compute.

The contribution of a particular graph $G$ to the free energy can be expressed
in the form
\be
T_r (G) = C_G (t/J)^r z_0^{-p} \sum_{s,l,m} a_{s,l,m} K^l e^{mK} \zeta^s
\ee
where $C_G$ is the embedding factor, or ``weak lattice constant" of
graph $G$ in the particular lattice considered, $p$ is the number
of points or vertices in the graph, the $a_{s,l,m}$ are numerical
constants, and the sum contains, for each graph, a finite
set of terms labeled by integers
$s,l,m$. Evaluation of these expressions is a lengthy procedure,
involving a trace over a space of $8^p$ states and evaluation, for
each term in the trace, of an $r$-fold multiple integral.
It is possible to find many time saving refinements, but, even so,
the evaluation of the worst case, the octagon, took some 260 hours
of CPU time on a  1GHz Compaq  alpha processor.

Having computed the $T(G)$ factors for all graphs it is then a simple
matter to combine these and to obtain, for any lattice, the free energy per site
in the form
\be
-\beta f = \ln z_0 + \sum_{r=2}^{\infty} z_0^{-r} F_r (K, \zeta) (t/J)^r \label{eq15}
\ee
where the $F_r$ are complete expressions in $K$ and $\zeta$, of
the form
\be
F_r (K , \zeta) =\sum_{s,l,m} a_{s,l,m} K^l e^{mK}\zeta^s \label{eq16}
\ee
These expressions are too lengthy to display here, but can be supplied
on request. To give some idea of the size, the
8th order factor $F_8$ contains 1042 separate terms.

From (\ref{eq15}) and (\ref{eq16}) one can compute expressions for the
internal energy and specific heat. 
The internal energy can be expressed in the form
\bea
u &=& - {\partial \over \partial \beta}{\Big \vert}_{\zeta} (-\beta f) \nonumber \\
  &=& u_0 + \sum_{r=2}^{\infty} z_0^{-(r+1)} E_r (K,\zeta ) (t/J)^r
\eea
where $u_0$ is the atomic limit result (\ref{eq3}) and the
$E_r(K,\zeta) $ are complete expressions. The specific heat is
$C_v=du/dT$.
For most purposes it is more
useful to express the series in terms of electron density $n$,
which can be obtained from (\ref{eq15}) via
\bea
n &=& \zeta {\partial \over \partial \zeta} (-\beta f) \nonumber \\
&=& n_0 + \sum_{r=2}^{\infty} z_0^{-(r+1)} Y_r(K,\zeta) (t/J)^r
\eea
where $n_0$ is the atomic limit result (\ref{eq4}) and the $Y_r(K,\zeta)$
are, again, complete expressions in $K$, $\zeta$. For fixed $n$ and
$K$ we then use a numerical reversion procedure to obtain
expansions for the fugacity, in power of $t/J$, which can then
be eliminated from the thermodynamic functions. For half-filling
$(n=1)$ this reversion can be carried out analytically.

In addition we have included a magnetic field term
\be
H' = H - h\sum (n_{i,\uparrow} - n_{i,\downarrow} + 2 S_i^z)
\ee
to allow calculation of the zero-field magnetic susceptibility $\chi_s$,
which is expressed in the form
\be
\beta^{-1} \chi_s = \beta^{-1} \chi_0 + \sum_{r=2}^{\infty} 
z_0^{-(r+1)} X_r (K, \zeta) (t/J)^r \label{eqchi}
\ee
where, again, the first term is the atomic limit result (\ref{eq7})
and the $X_r$ are complete expressions. The fugacity can again be
eliminated in favour of the electron density, as described above.

Several checks on the correctness of our results have been made.
Since temperature enters explicitly, we can take the zero temperature
($K\to\infty$) limit analytically, to recover the ground state
energy series\cite{zwh,shi95}. Another test is the $J\to 0$ limit,
in which $F_r(K,\zeta)$ in (\ref{eq15}) has a leading term of
order $K^r$, giving an expression in powers of  $(\beta t)$. This can be
compared with the results for free electrons.
Complete agreement is obtained. We are confident that our
results are correct, but cannot exclude the possibility of errors
not picked up by these checks.

The analysis of the results follows standard lines.
For any fixed $K,\zeta$ we obtain expressions for thermodynamic
quantities as a series in the single variable $(t/J)$. For
small $t/J$ the series converges rapidly and even a naive sum
gives good accuracy.
Pad\'e approximants and integrated differential approximants\cite{gut}
allow extrapolation to larger $(t/J)$. For fixed $n$ we obtain series
for $\zeta$ in terms of $t/J$, which are then substituted into
corresponding expressions (\ref{eq16}), (\ref{eqchi}) to obtain a single-variable
series in $(t/J)$, which is then evaluated as before. To calculate the
specific heat it is necessary to include derivatives of $\zeta$ with respect
to temperature. For the benefit of the reader we give in Table I coefficients
for some representative cases. Other series can be supplied on request.

\begin{figure}[t]
  { \centering
    \includegraphics[width=\figwidths]{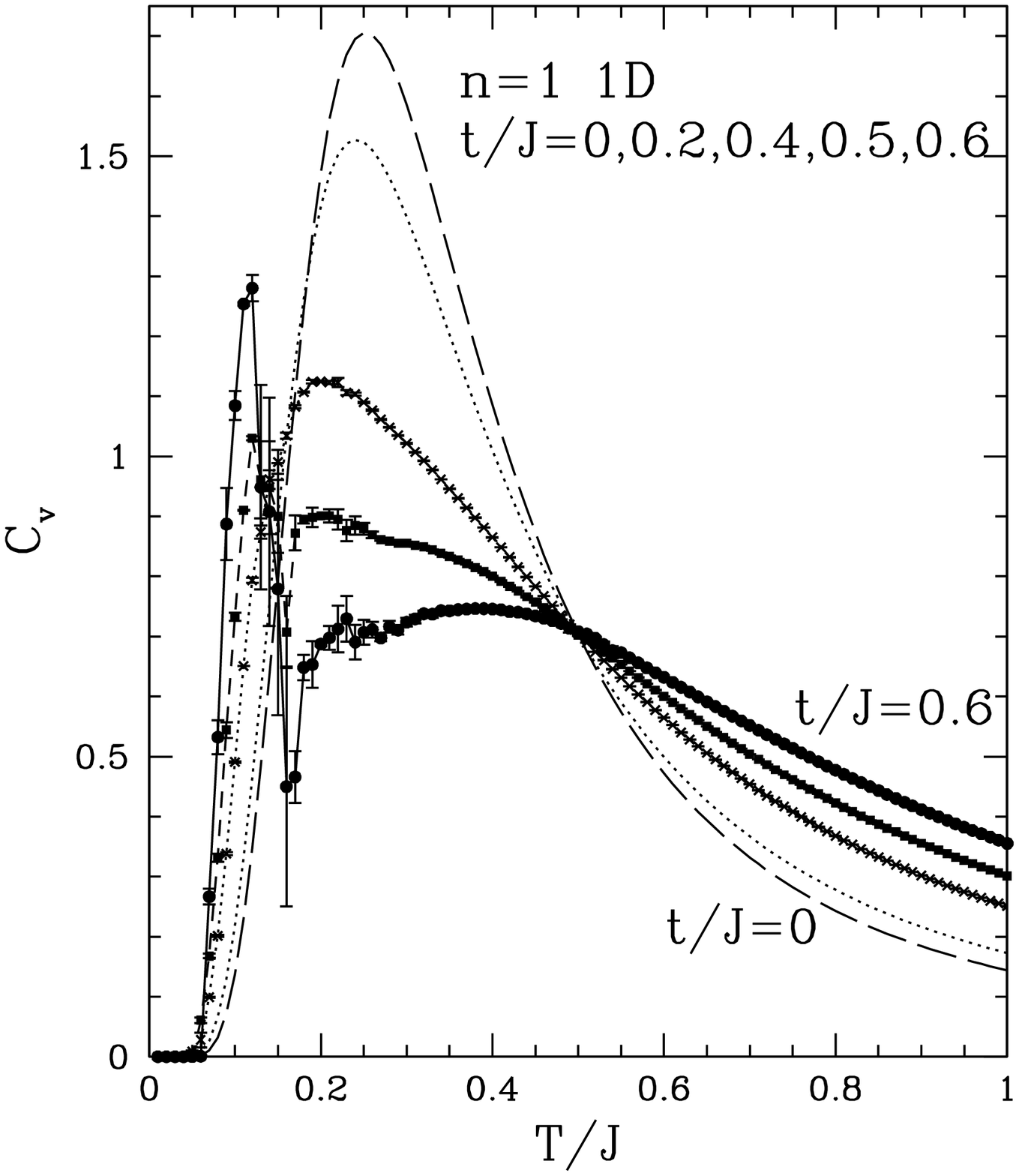}  \hspace{-0.75cm}
    \includegraphics[width=\figwidths]{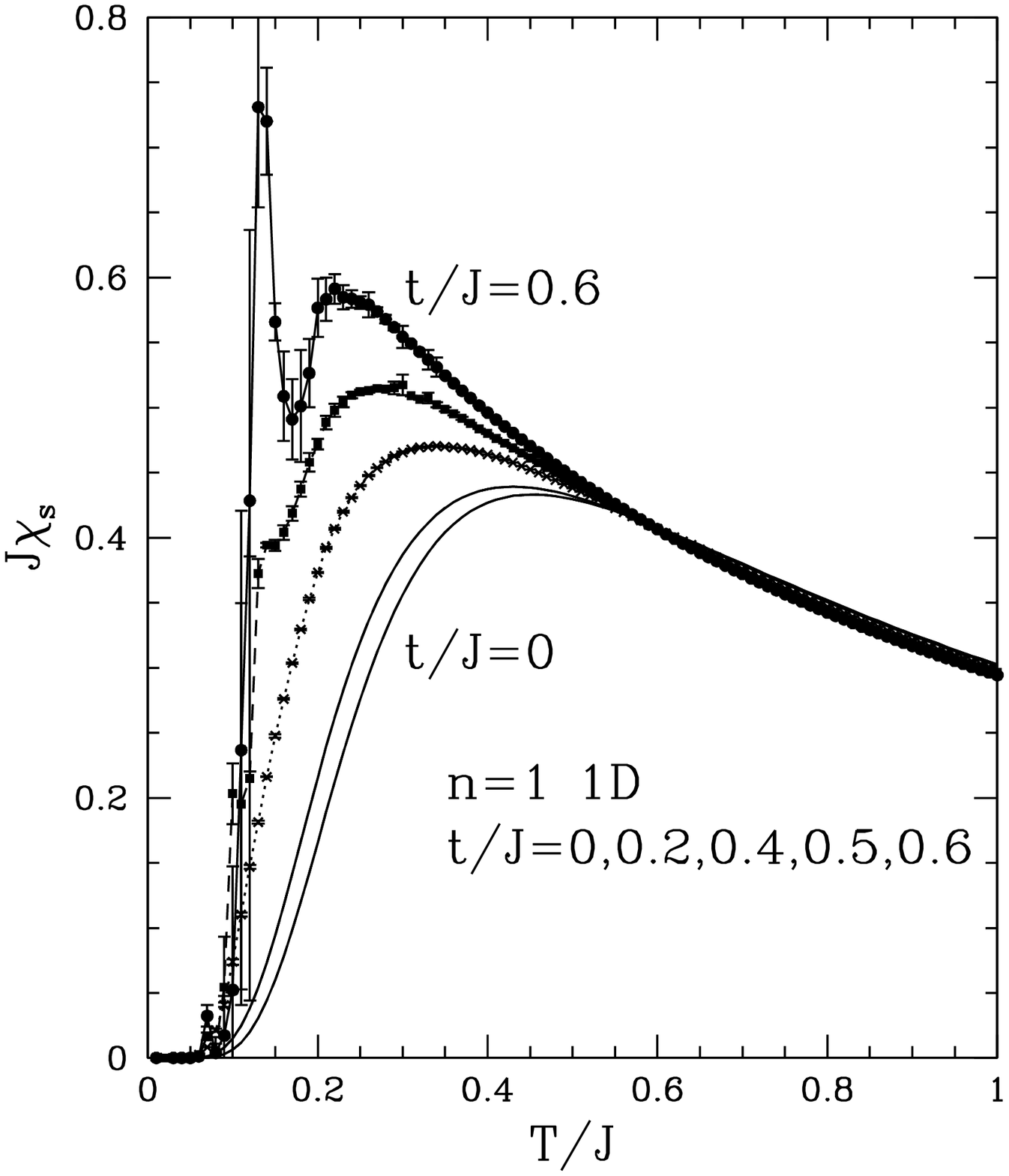}\hspace{-0.75cm}
    \includegraphics[width=\figwidths]{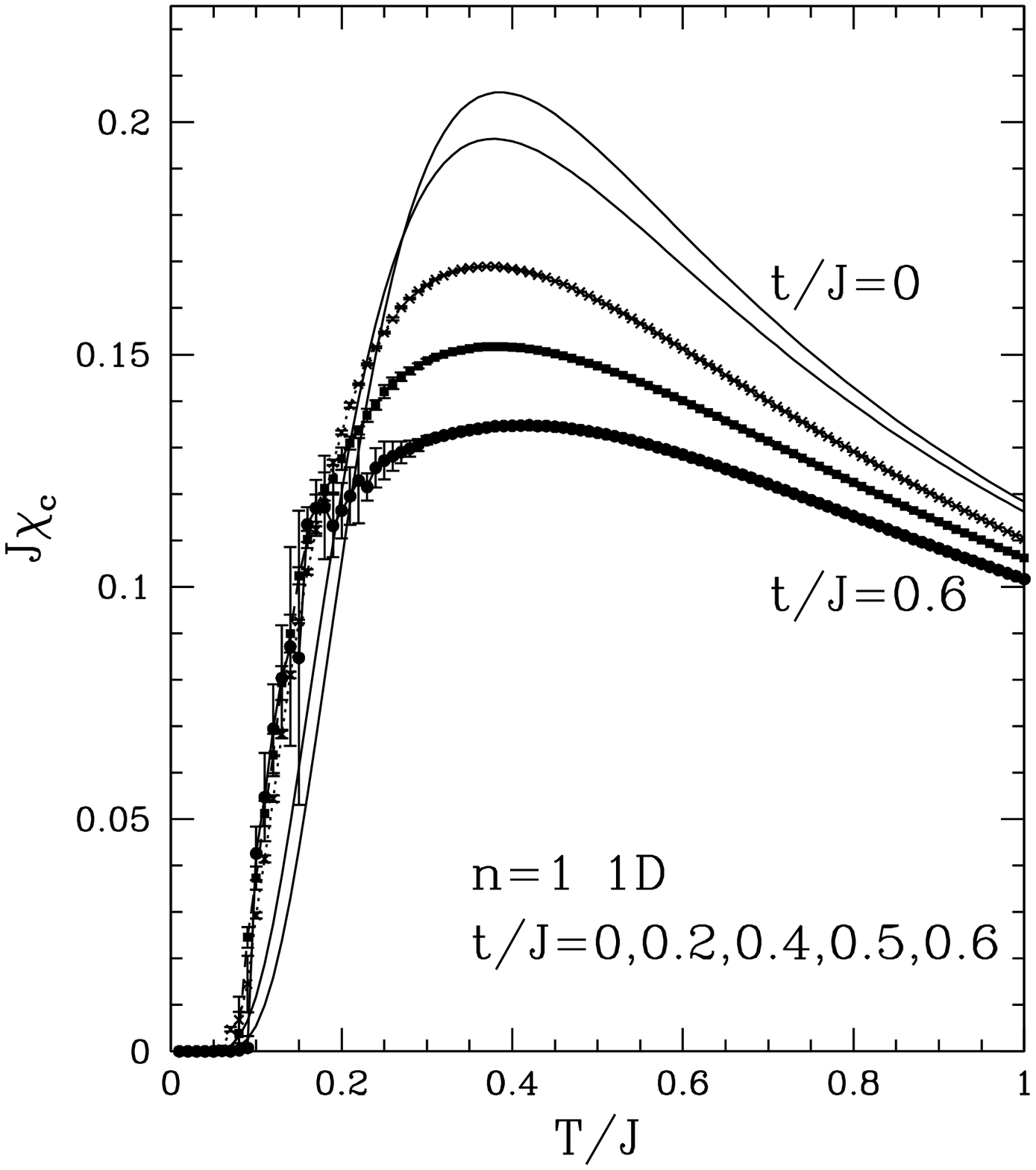}
    \caption{The specific heat $C_v$ (a), magnetic susceptibility $\chi_s$ (b)
    and charge susceptibility $\chi_c$ (c) versus $T/J$ for the linear chain at
    $n=1$ for $t/J=0,0.2,0.4,0.5,0.6$.
    \label{fig_1D_n=1}}
  }
\end{figure} 

\begin{figure}[b]
  { \centering
    \includegraphics[width=\figwidths]{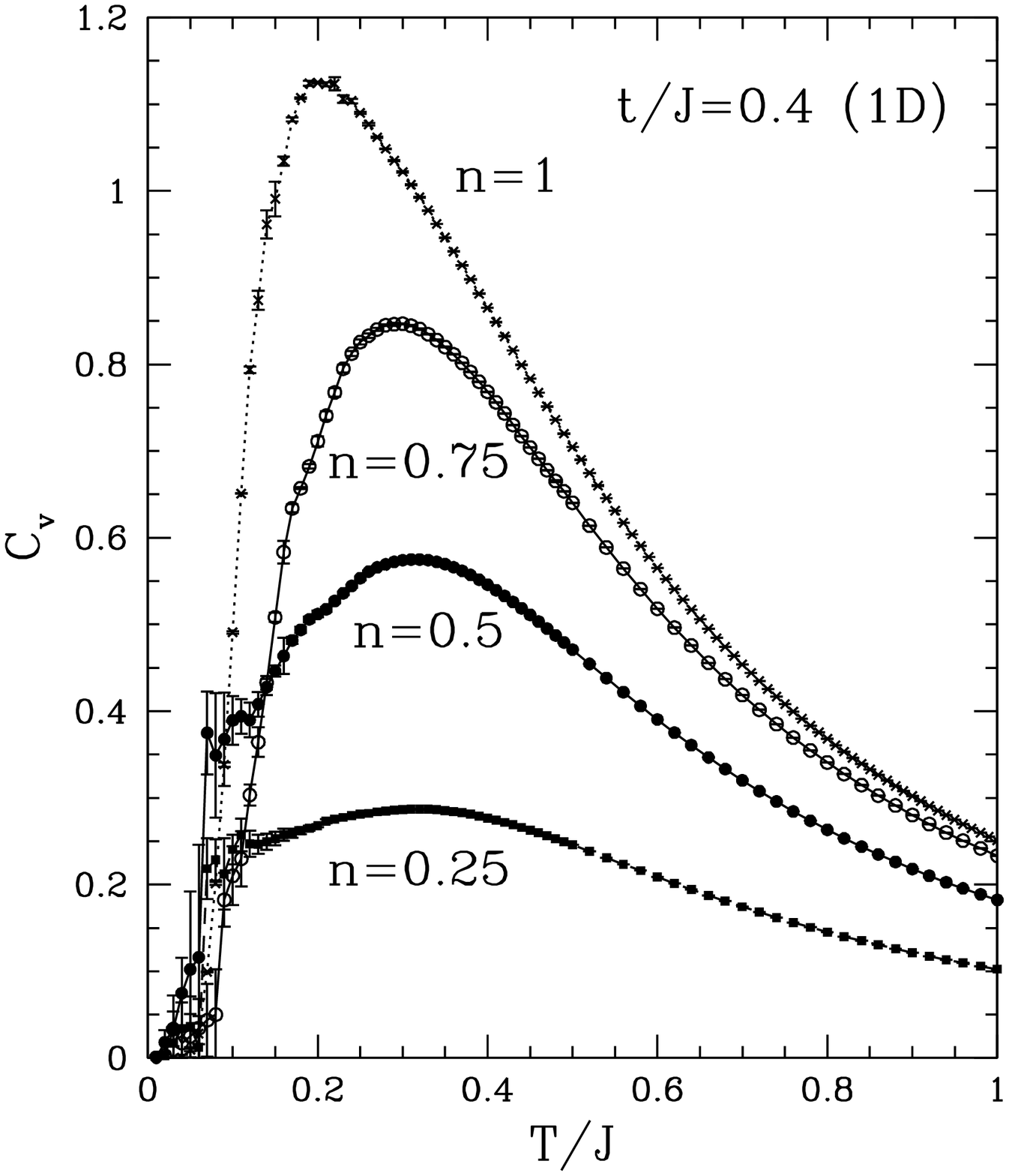}  \hspace{-0.75cm}
    \includegraphics[width=\figwidths]{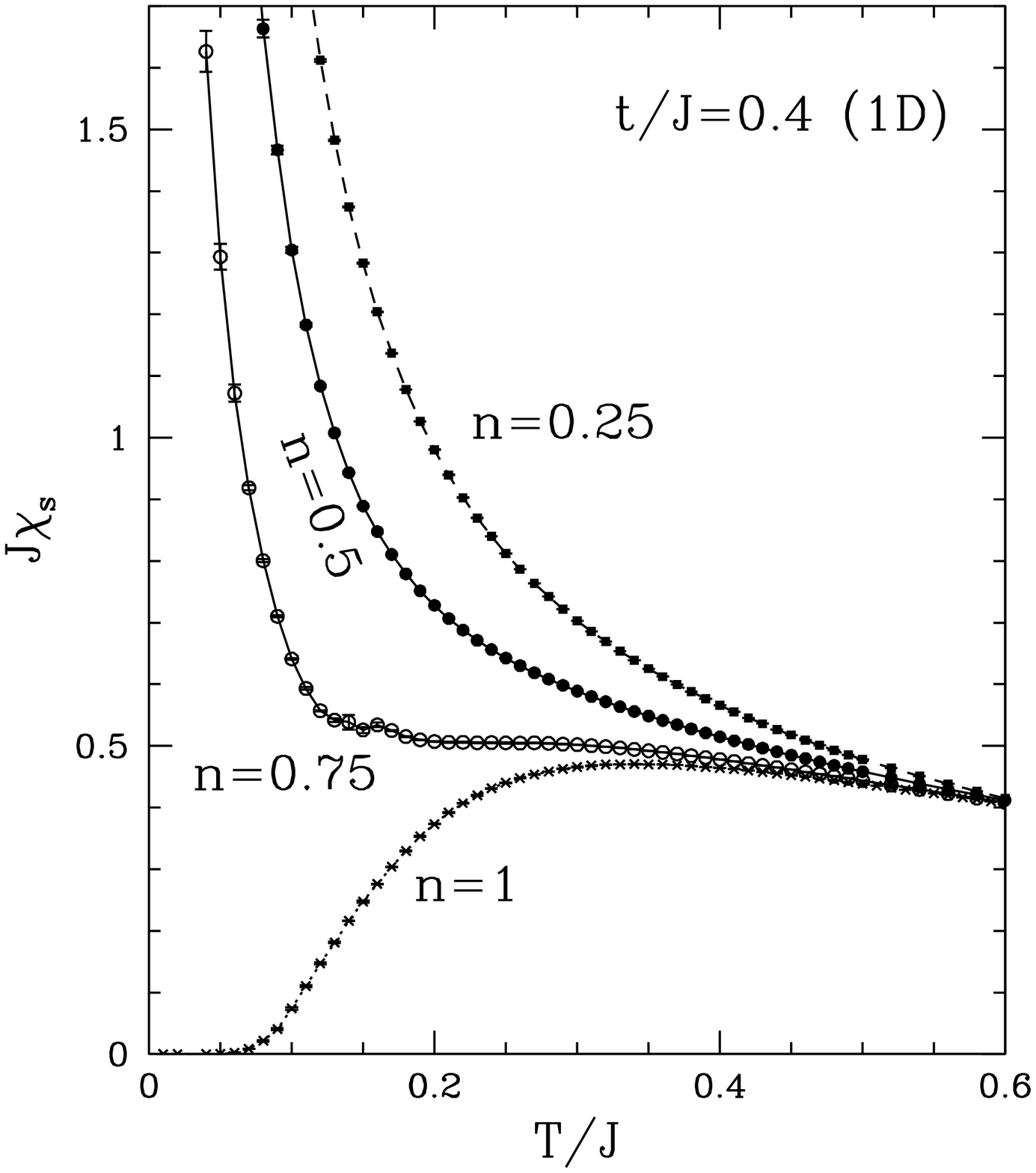}\hspace{-0.75cm}
    \includegraphics[width=\figwidths]{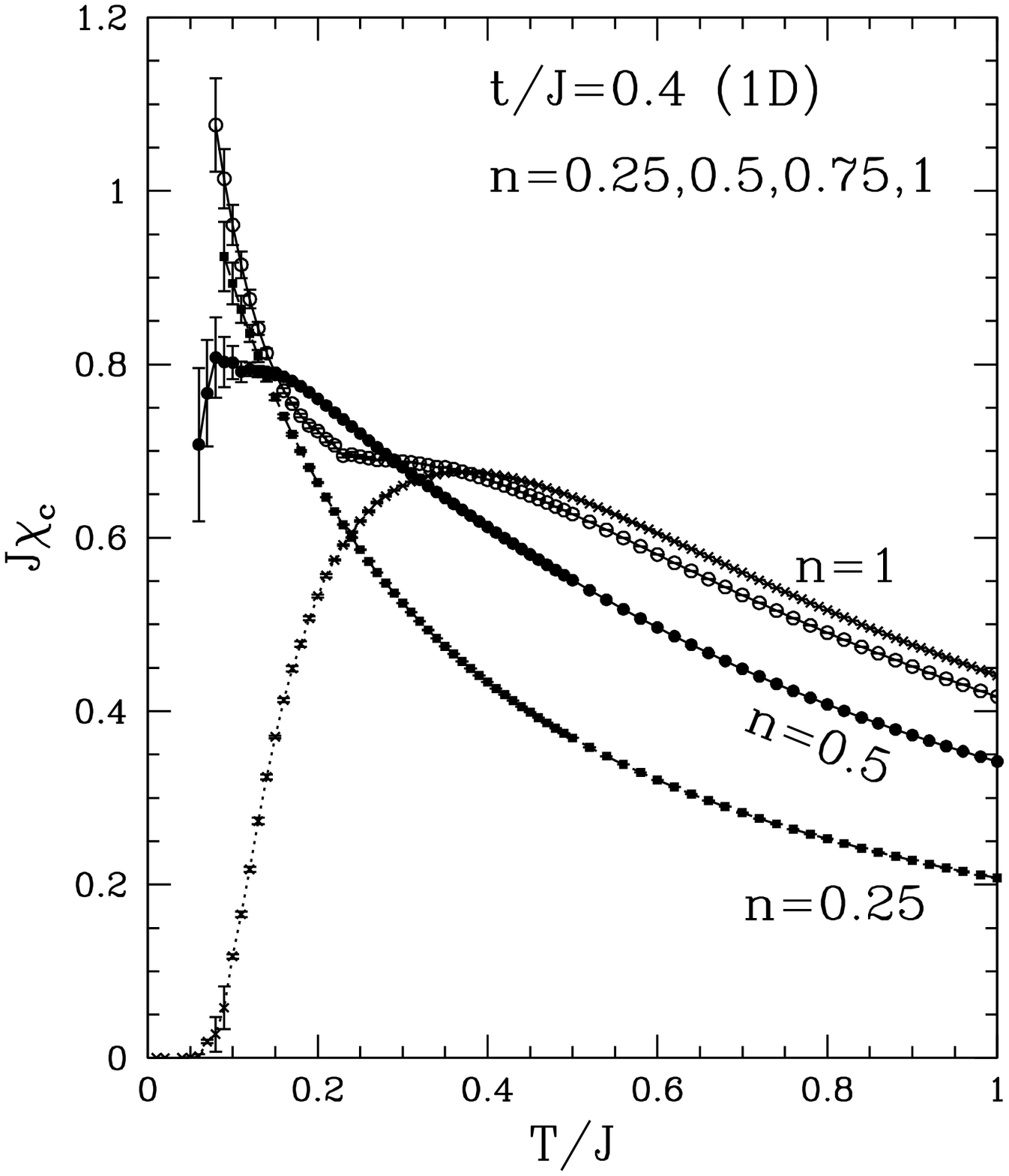}
    \caption{The same as Fig.1, but for $t/J=0.4$ and $n=1,0.75,0.5,0.25$.
    \label{fig_1D_n}}
  }
\end{figure} 

While the general expressions are too lengthy to write down,
at high temperatures, one can expand the various quantities
in powers of $\beta$ or $K$. The expressions up to order 
$K^4$ are provided in the Appendix. These are valid
for all loosed-packed/bipartite lattices.

In the following sections we present results for the linear
chain, the square lattice (sq) and the simple cubic (sc) lattice.
A discussion of other lattices will be presented elsewhere.

\section{\label{sec:1d}The 1D Kondo Lattice Model}

We consider the antiferromagnetic model on a linear chain, with
arbitrary electron concentration $n$. Using  the procedure described
above we have computed the specific heat and spin and charge susceptibilities. 
Figure 1(a), (b) (c) shows these quantities, as functions
of temperature, for $n=1$ (half-filling) for $t/J = 0, 0.2, 0.4, 0.5,0.6$.
For $t/J \lesssim 0.4$ the series are well converged and the curves are
obtained simply from the partial sums. For the larger values integrated
differential approximants have been used, with the error bars indicating
the variation between different approximants. Good agreement is obtained
with the DMRG results for all three quantities. However our (relatively
short) series are unable to probe the larger $t/J$  region as well
as DMRG can.

\begin{figure}[t]
  { \centering
    \includegraphics[width=\figwidth]{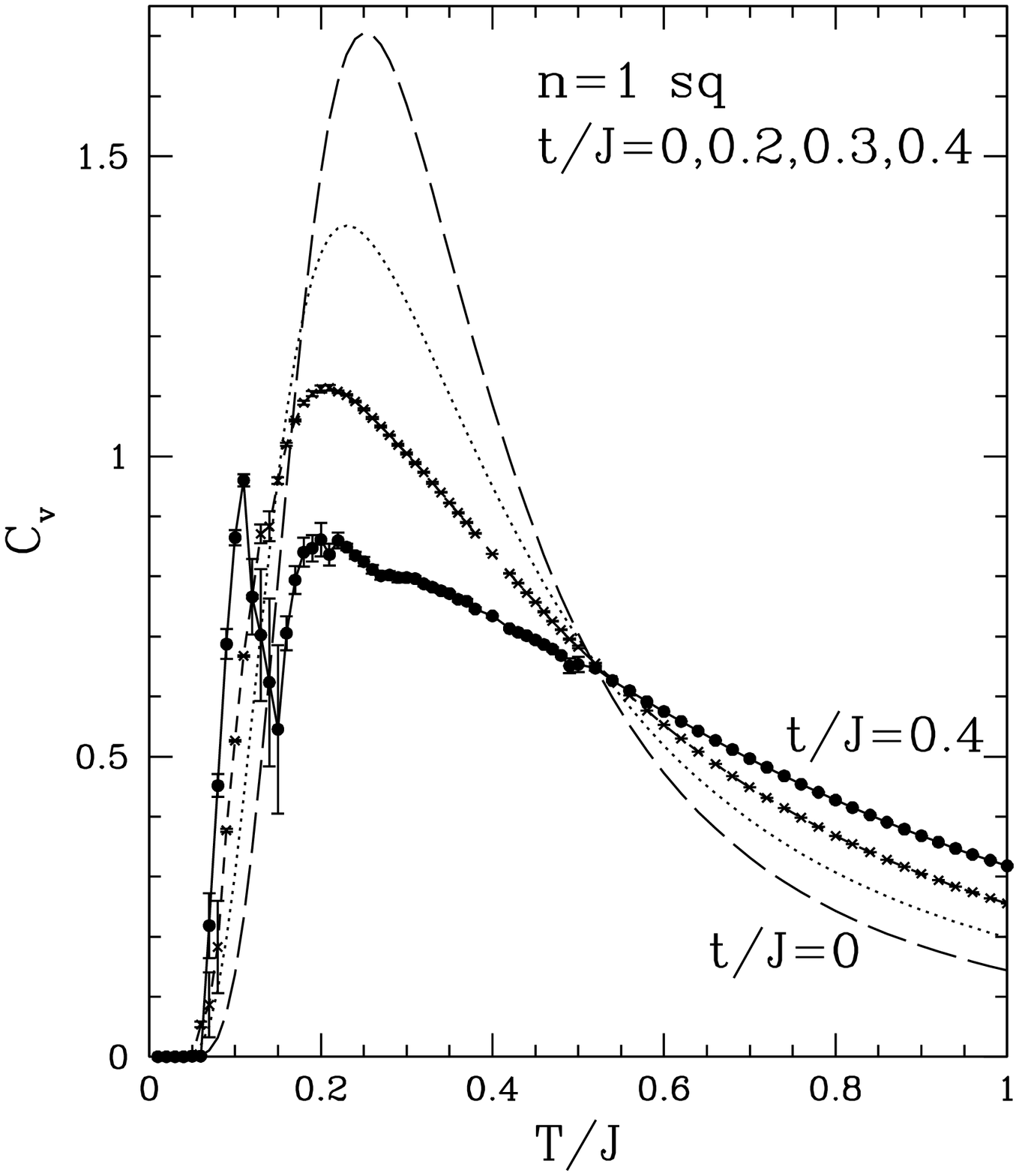}  \hspace{1cm}
    \includegraphics[width=\figwidth]{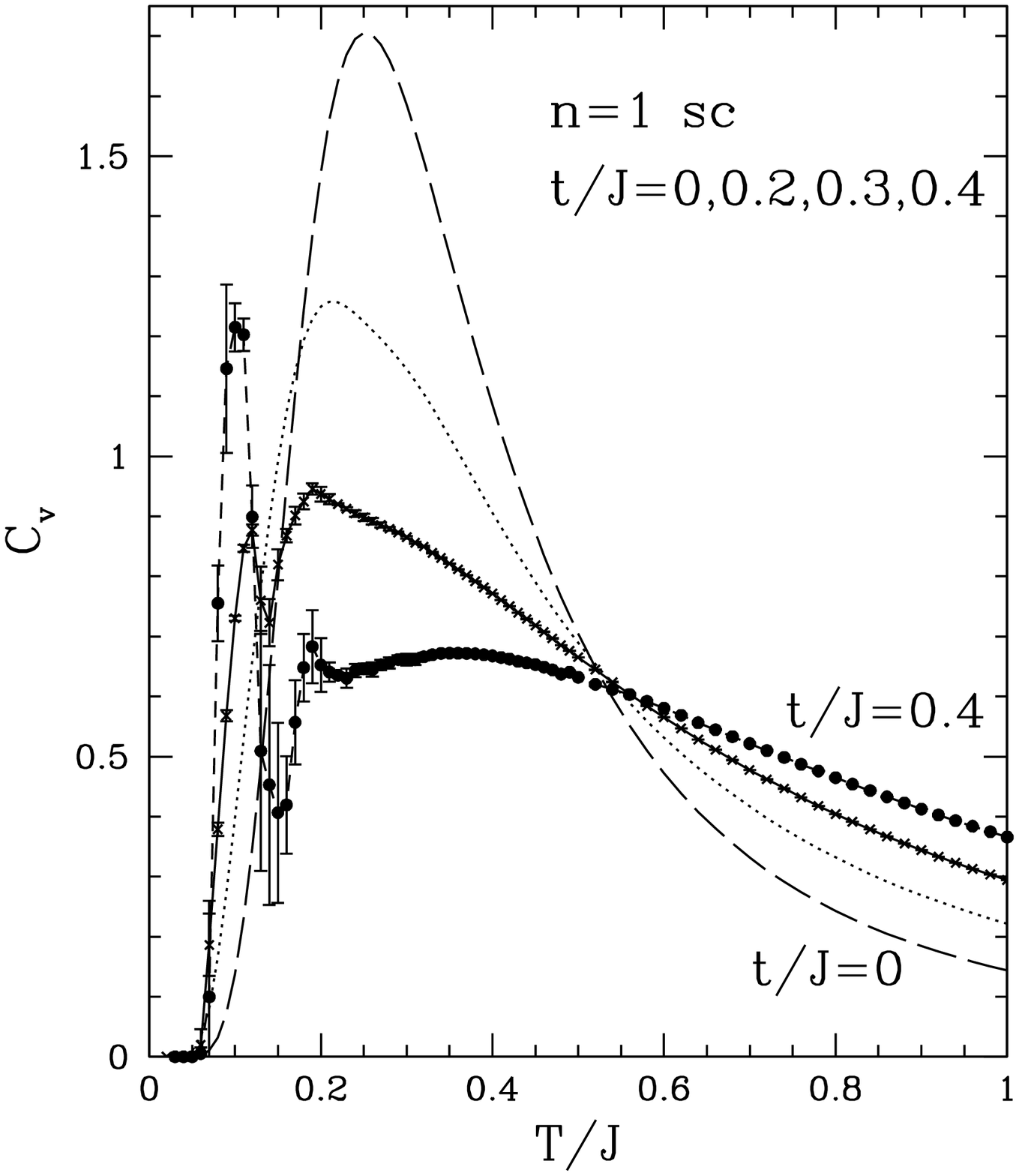}
    \caption{The specific heat $C_v$  versus $T/J$ for square lattice (a) and
    simple cubic lattice (b) at
    $n=1$ for $t/J=0,0.2,0.3,0.4$.
    \label{fig_23d_Cv_n=1}}
  }
\end{figure} 

\begin{figure}[b]
  { \centering
    \includegraphics[width=\figwidth]{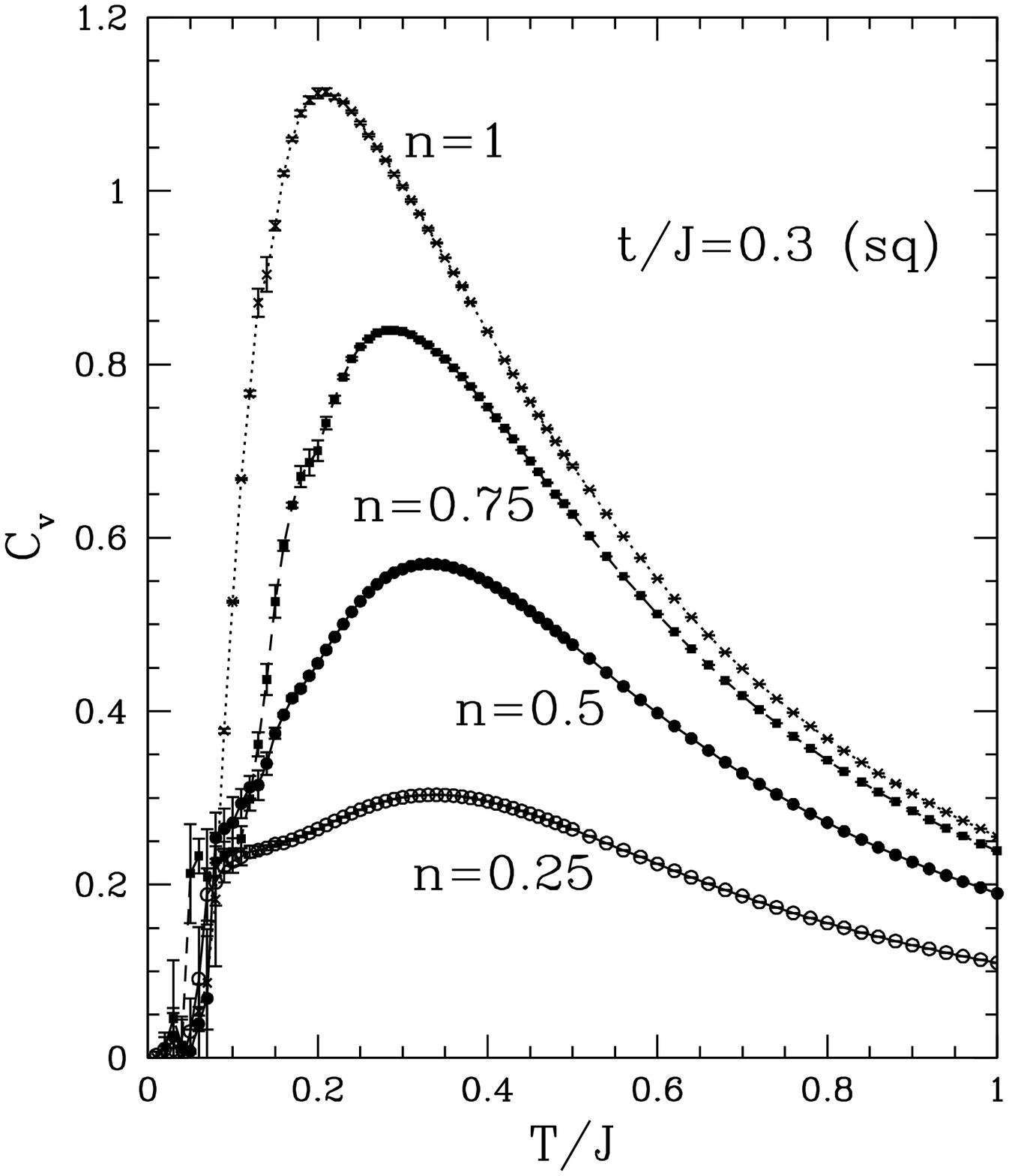}  \hspace{1cm}
    \includegraphics[width=\figwidth]{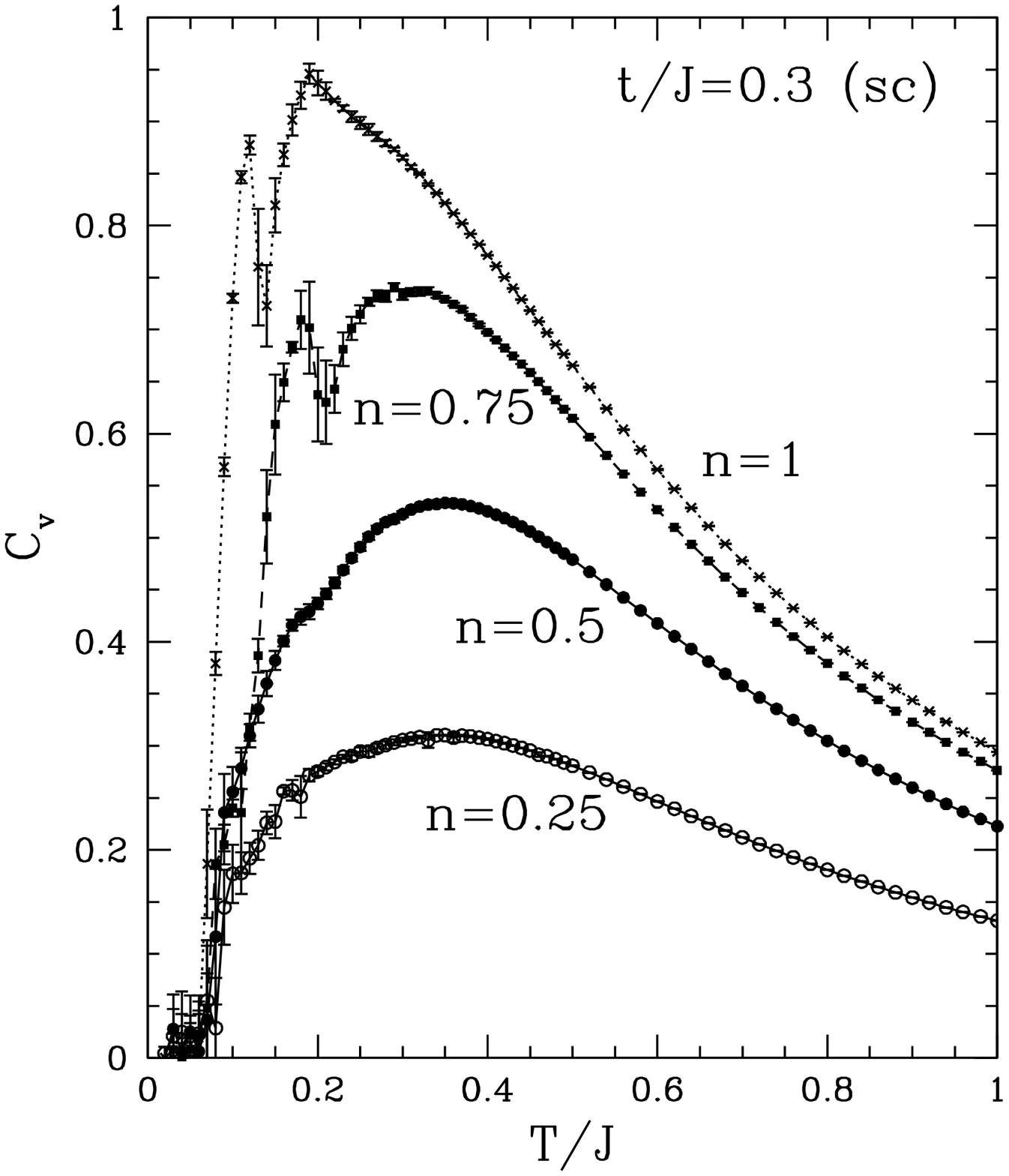}
    \caption{The specific heat $C_v$  versus $T/J$ for square lattice (a) and
    simple cubic lattice (b) at
    $t/J=0.3$ for $n=1,0.75,0.5,0.25$.
    \label{fig_23d_Cv_n}}
  }
\end{figure} 

Some comments on Fig. 1 are in order. Fig. 1(a) 
shows the specific heat, which shows an interesting crossover 
from a single peak for small $t$ to a two-peak structure at larger $t$. 
In the larger $t$ region the high temperature peak becomes
broadened and less prominent and is, presumably, due to conduction electrons
whereas the low-temperature peak arises from the fluctuating local
spins. The spin susceptibility (Fig. 1(b)) also has a peak at
a characteristic temperature. 
The peak is enhanced and moves to lower $T$
on increasing the hopping parameter $t$. Increasing $t/J$
will weaken the singlet correlations and a lower characteristic 
temperature is sufficient for thermal fluctuations to become
dominant. There is some indication of a double peak for
$t/J=0.6$, but this may be an artifact of the numerical
analysis. The charge susceptibility also has peak at a characteristic temperature,
and a rapid drop to zero at low temperatures.
The peak is depressed with increasing $t$ but the position stays relatively
constant.

To display the effect of varying conduction electron density we have
chosen an intermediate value $t/J=0.4$ and show,
in Fig. 2(a), (b), (c) curves of $C_v$, $\chi_s$ and $\chi_c$ 
 versus temperature for $n=1$, 0.75, 0.5 and 0.25.
The high-temperature peak in
$C_v$ (Fig. 2(a)) drops roughly proportionally to
$n$, in agreement with the assignment of this peak to conduction electrons.
The series do not allow the low-T specific heat to be determined
with sufficient precision to see the effects of doping.
Tsunetsugu {\it et al.}\cite{tsu97} present a ground state phase
diagram of the 1D Kondo lattice (Fig. 6 of Ref. \onlinecite{tsu97}),
where a transition line separates a small $n$ ferromagnetic phase
from a large $n$ paramagnetic phase. For the parameter ratio $t/J=0.4$
the critical doping is $n_c \simeq 0.65$. While there cannot be true
order at finite temperature, the marked change in the low T specific
heat between the two curves with $n=0.75$ and 0.5 may will be a reflection of
this effect.
The effect of doping on $\chi_s$ is dramatic. For $n=1$, at low temperatures,
the system is in a gapped singlet phase and $\chi_s$ goes to zero
exponentially. Away from half-filling there will be free  spins 
and $\chi_s$ diverges according to the usual Curie law.
The charge susceptibility $\chi_c$ also shows a sharp crossover between
the undoped and doped cases. 

\begin{figure}[t]
  { \centering
    \includegraphics[width=\figwidth]{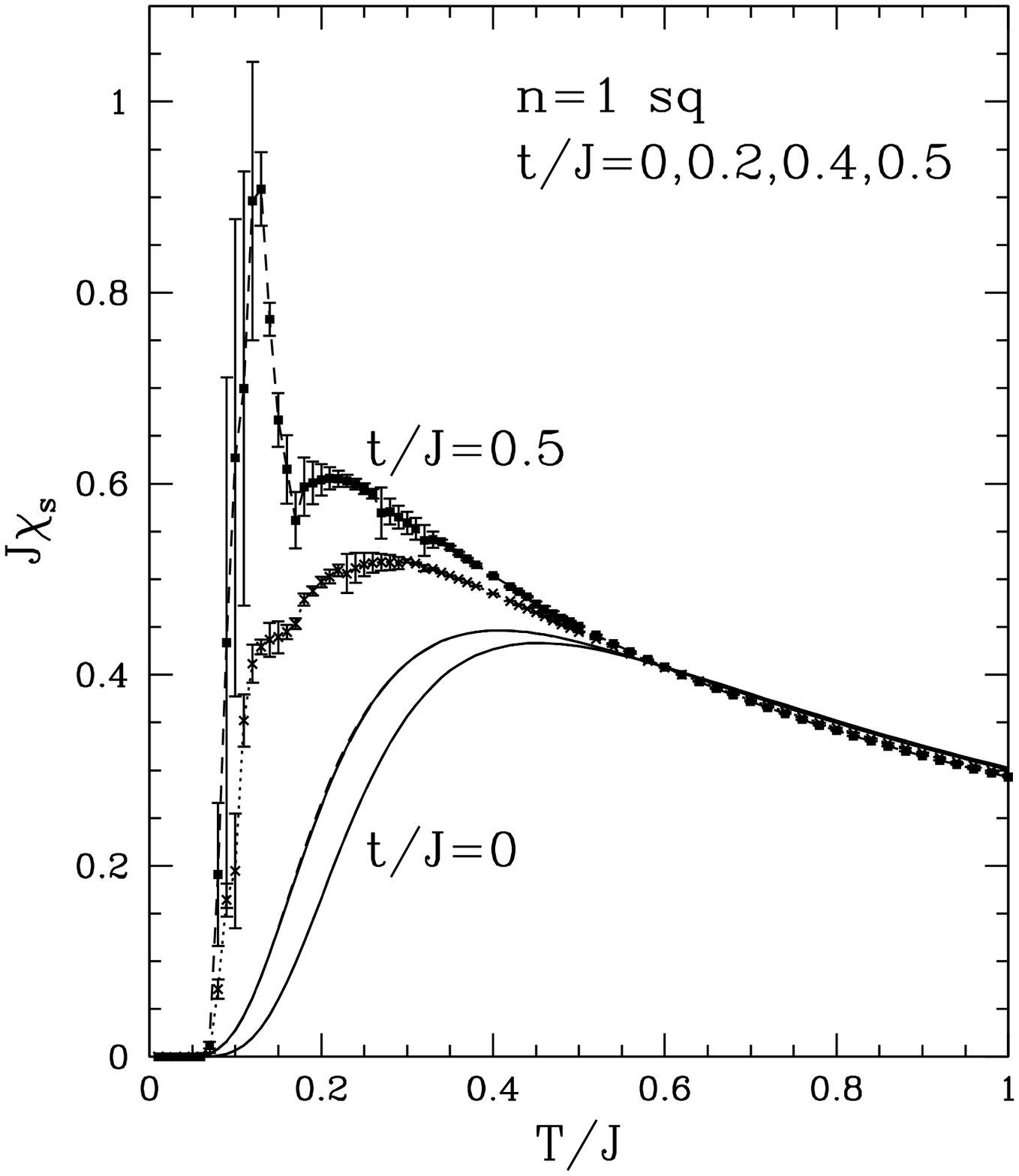}  \hspace{1cm}
    \includegraphics[width=\figwidth]{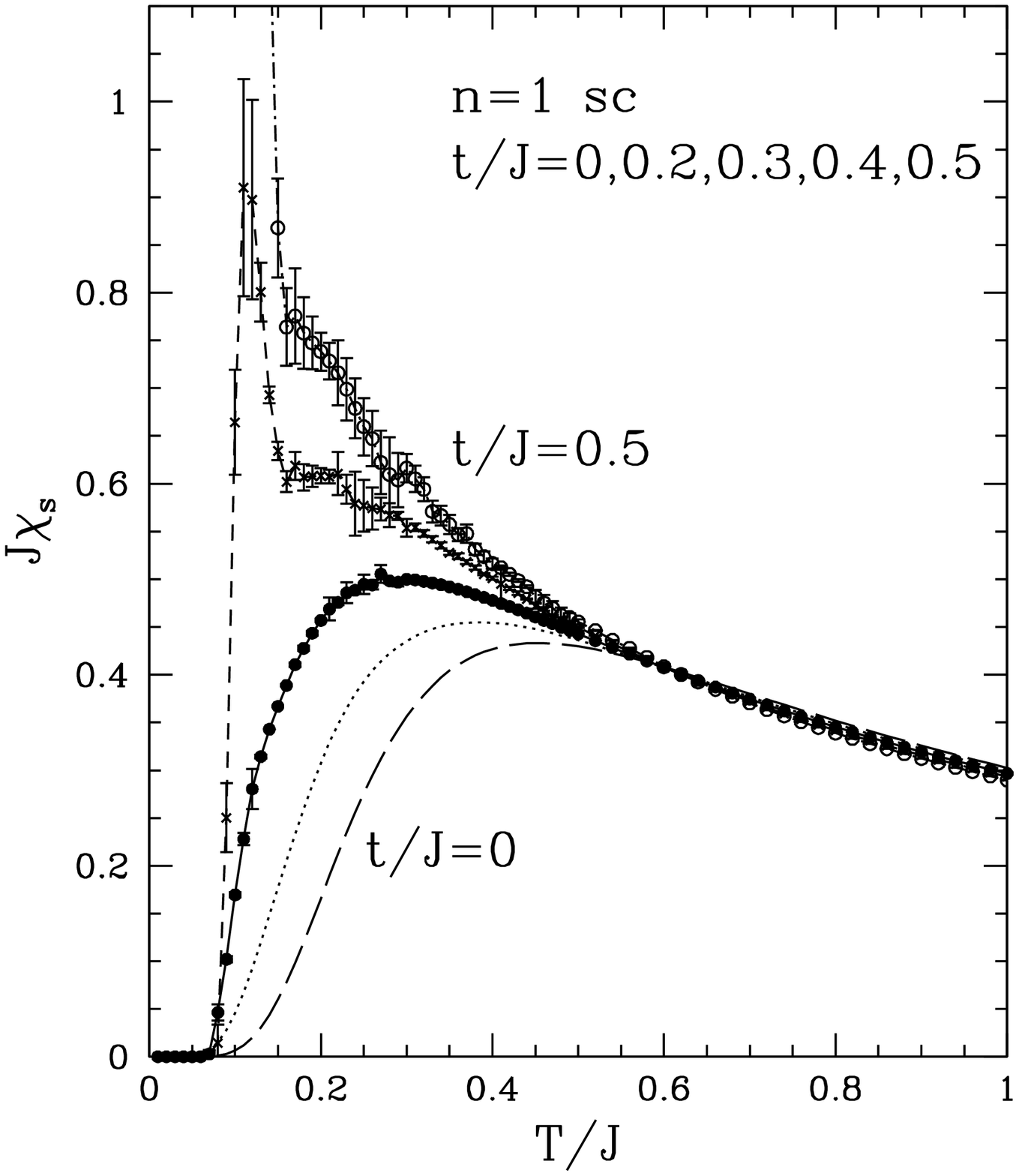}
    \caption{The magnetic susceptibility $\chi_s$  versus $T/J$ for square lattice (a) and
    simple cubic lattice (b) at
    $n=1$ for $t/J=0,0.2,0.3,0.4$.
    \label{fig_23d_chis_n=1}}
  }
\end{figure} 

\begin{figure}[b]
  { \centering
    \includegraphics[width=\figwidth]{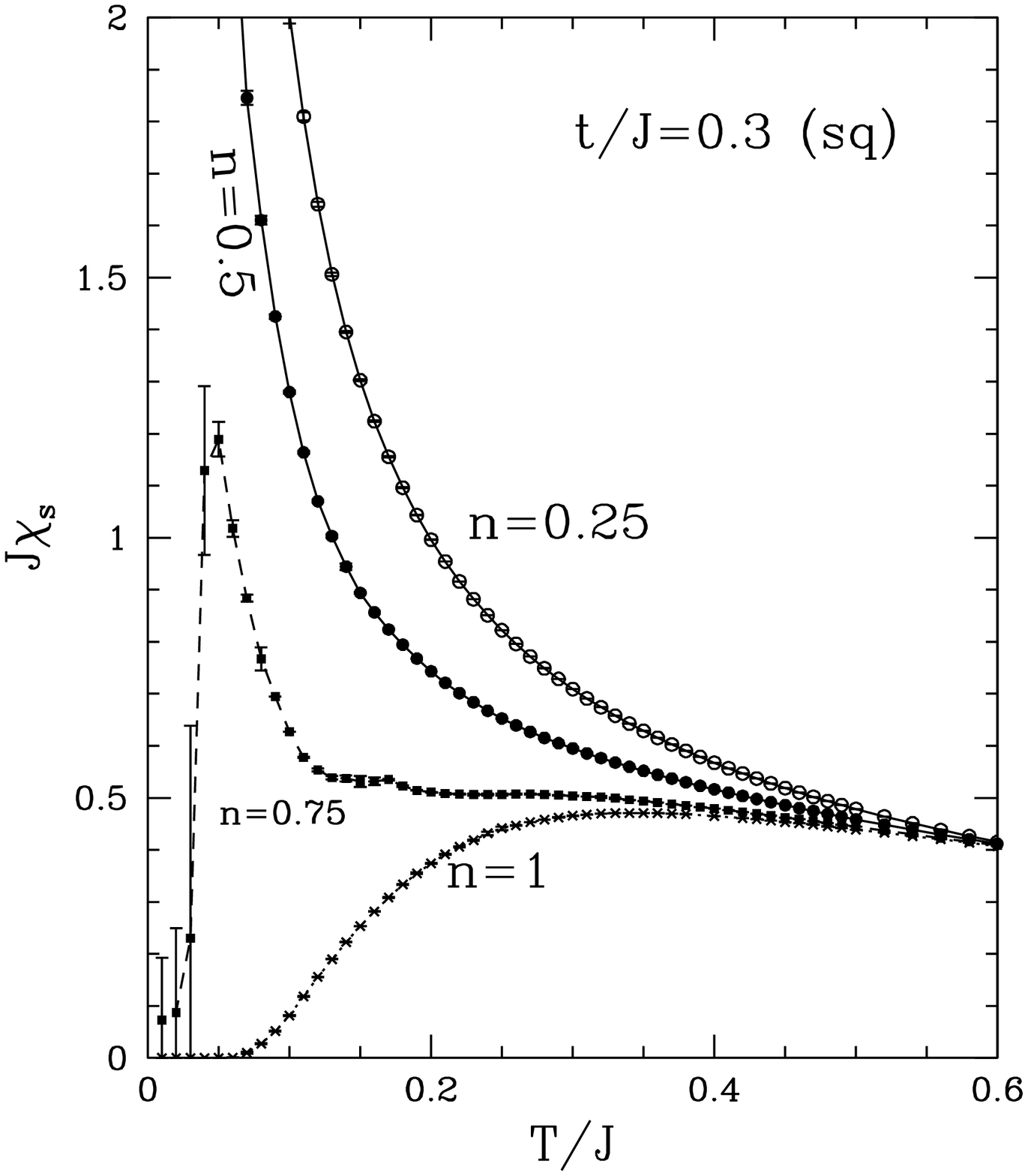}  \hspace{1cm}
    \includegraphics[width=\figwidth]{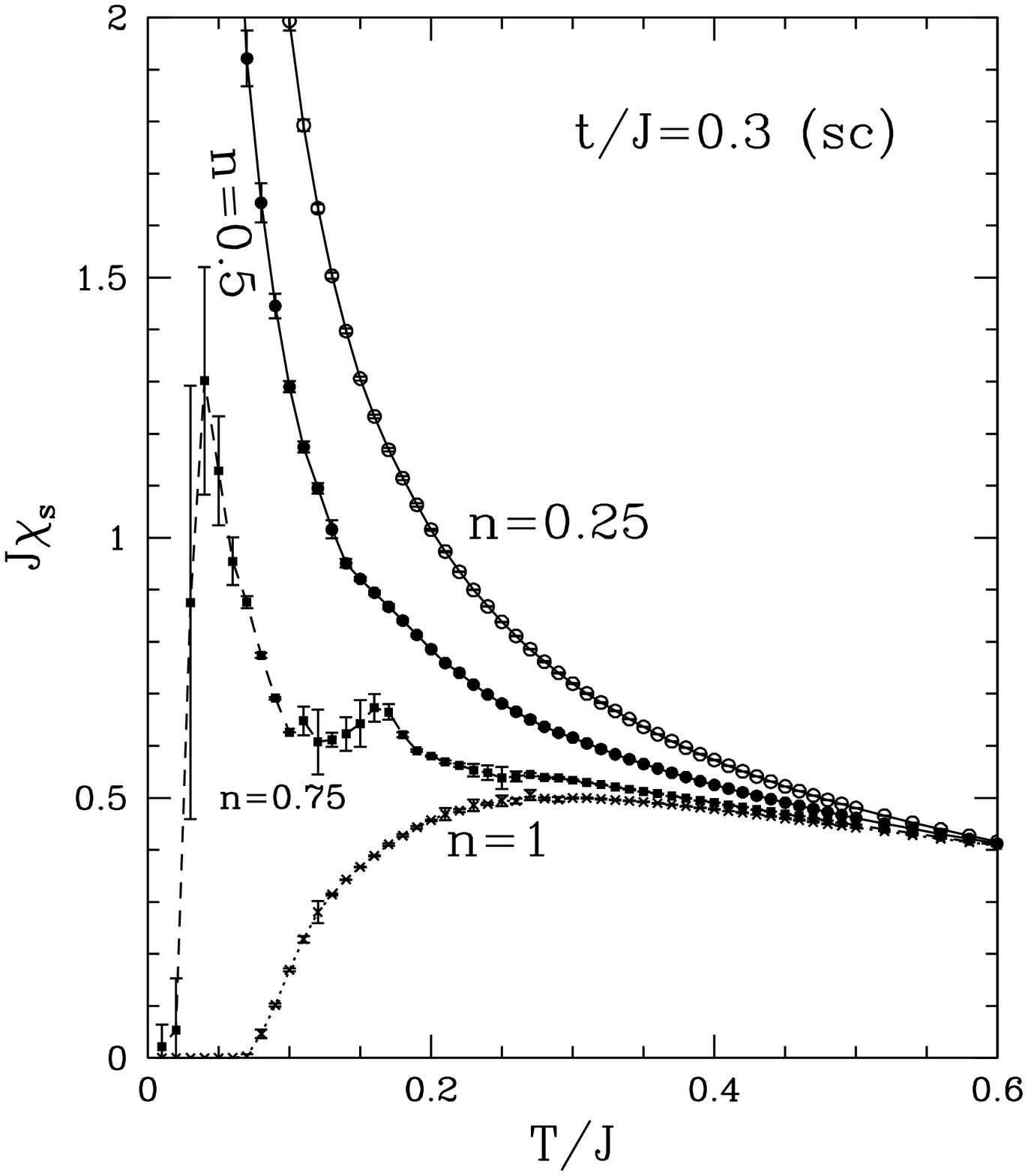}
    \caption{The  magnetic susceptibility $\chi_s$ versus $T/J$ for square lattice (a) and
    simple cubic lattice (b) at
    $t/J=0.3$ for $n=1,0.75,0.5,0.25$.
    \label{fig_23d_chis_n}}
  }
\end{figure} 

The 1D results are consistent with the,
presumably, more accurate DMRG calculations. 
We are not aware of any published DMRG results for specific heat at finite doping,
or for the susceptibility for $n<0.8$.
Our results  confirm that the
series approach can be successfully applied to this model. We now turn
to the 2D and 3D cases, where far less is known and where other methods 
have particular difficulties.

\begin{figure}[t]
  { \centering
    \includegraphics[width=\figwidth]{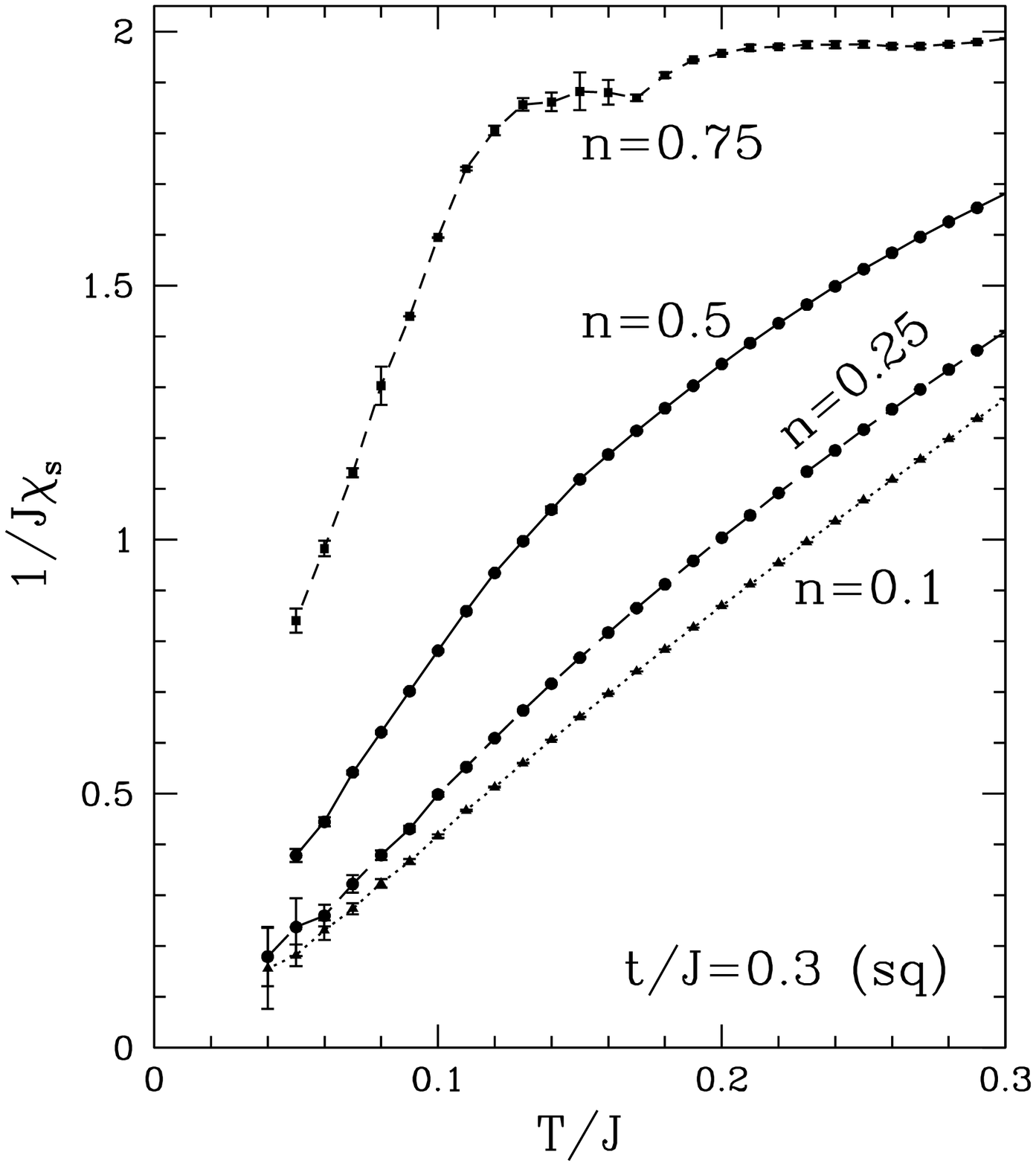}  \hspace{1cm}
    \includegraphics[width=\figwidth]{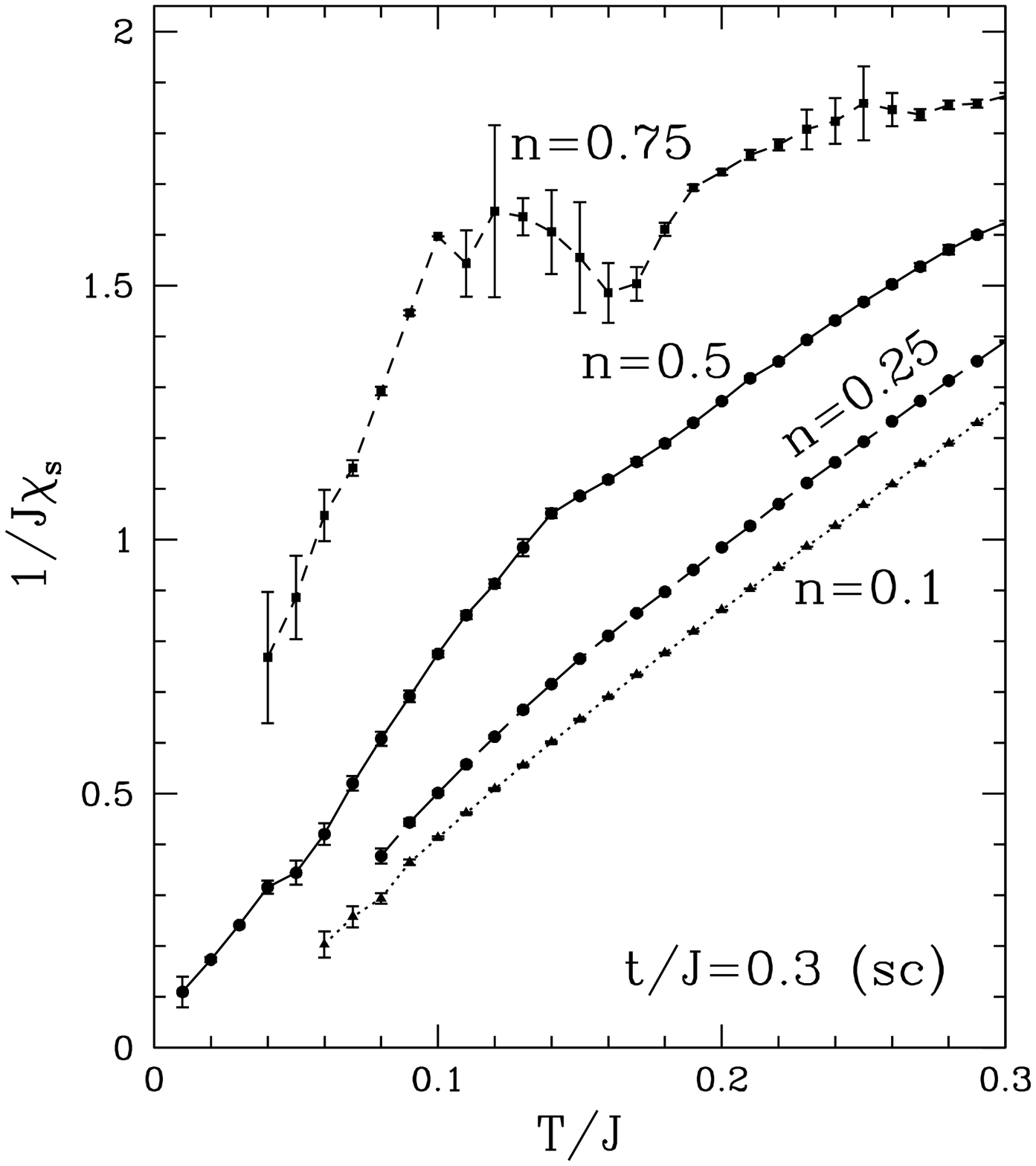}
    \caption{The inverse of magnetic susceptibility $1/\chi_s$ versus $T/J$ for 
    square lattice (a) and
    simple cubic lattice (b) at
    $t/J=0.3$ for $n=0.75,0.5,0.25,0.1$.
    \label{fig_23d_inv_chis_n}}
  }
\end{figure} 

\begin{figure}[b]
  { \centering
    \includegraphics[width=\figwidth]{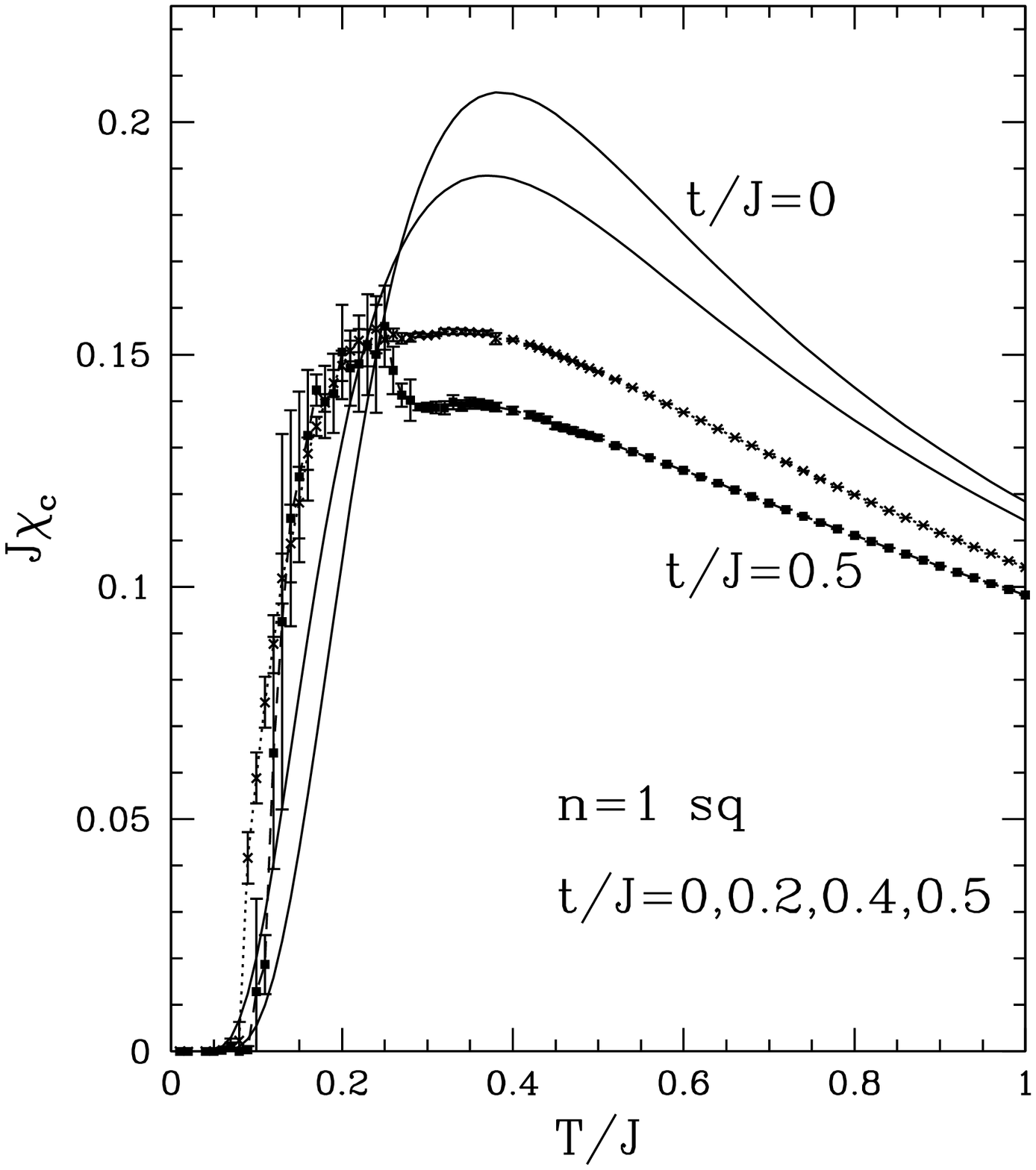}  \hspace{1cm}
    \includegraphics[width=\figwidth]{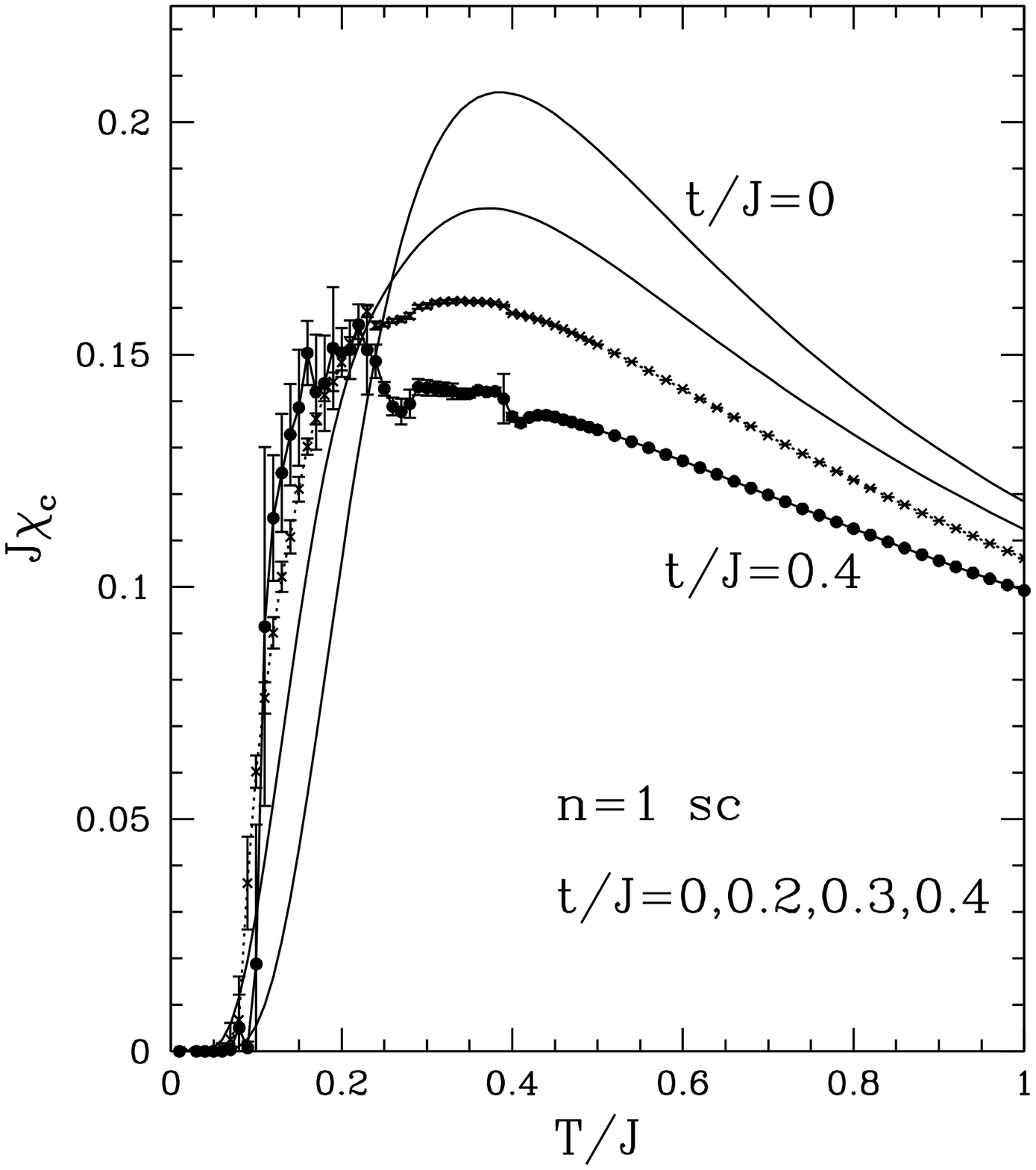}
    \caption{The charge susceptibility $\chi_c$ versus $T/J$ for square lattice (a) and
    simple cubic lattice (b) at
    $n=1$ for $t/J=0,0.2,0.3,0.4$.
    \label{fig_23d_chic_n=1}}
  }
\end{figure} 

\begin{figure}[t]
  { \centering
    \includegraphics[width=\figwidth]{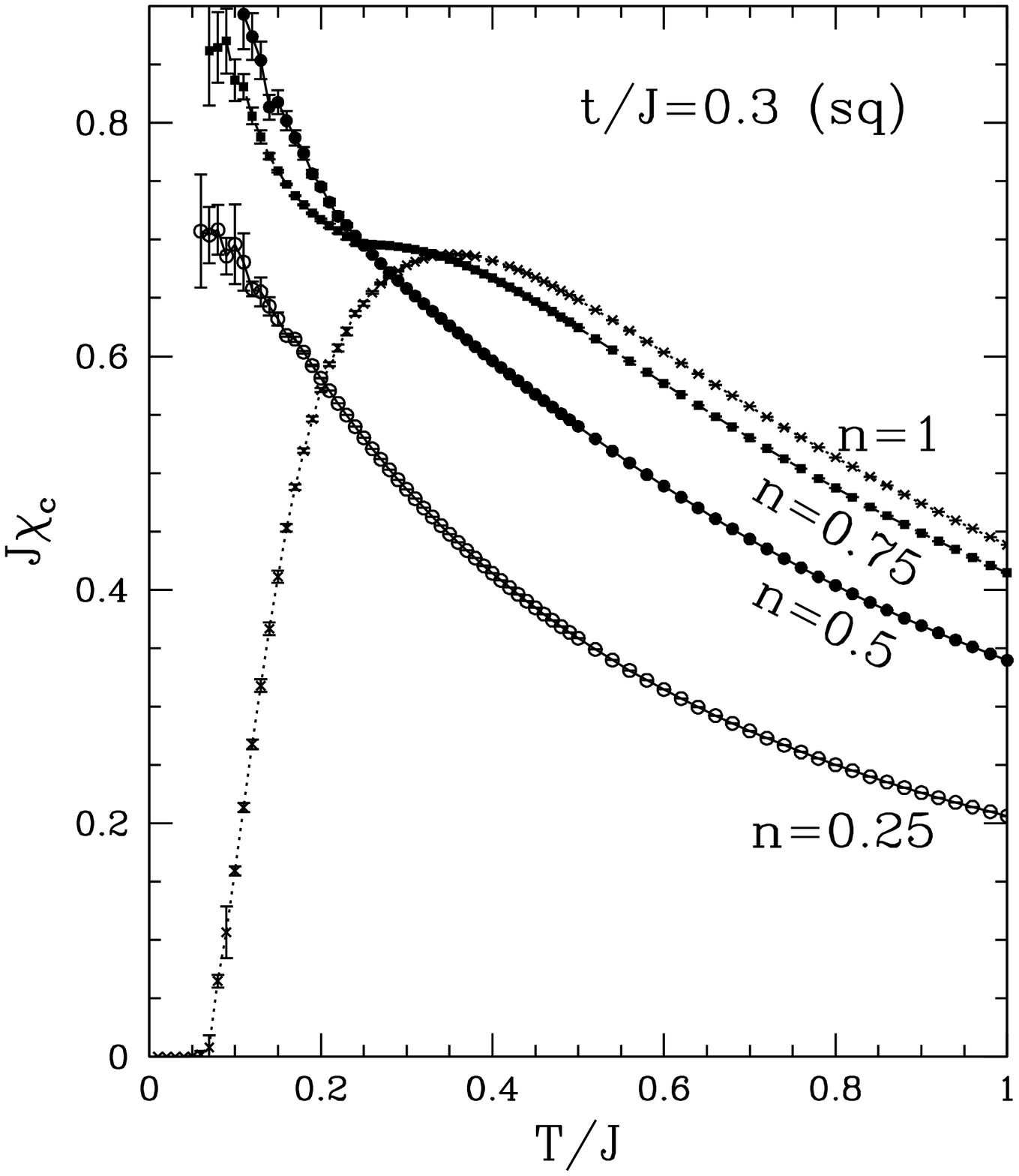}  \hspace{1cm}
    \includegraphics[width=\figwidth]{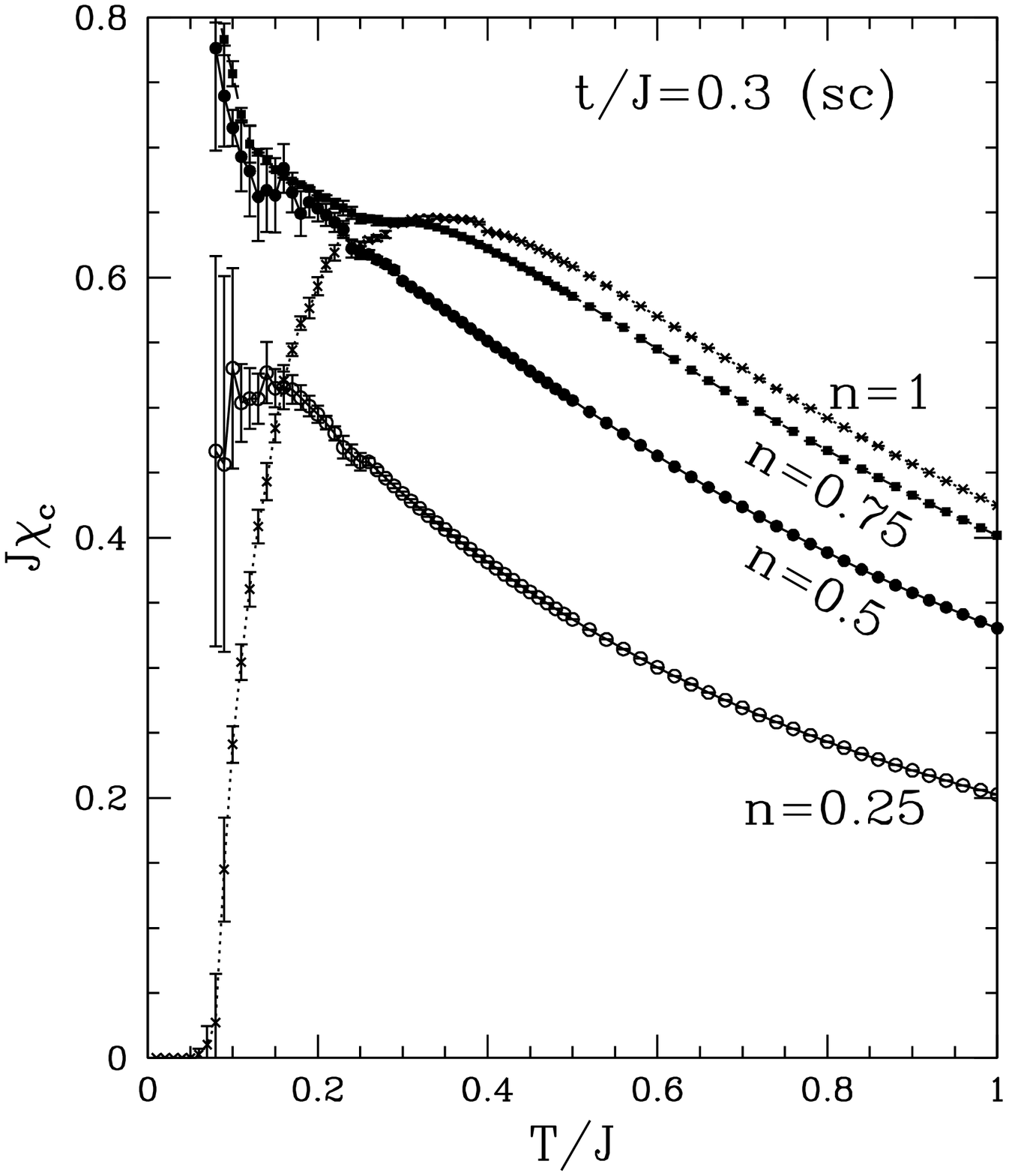}
    \caption{The charge susceptibility $\chi_c$ versus $T/J$ for square lattice (a) and
    simple cubic lattice (b) at
    $t/J=0.3$ for $n=1,0.75,0.5,0.25$.
    \label{fig_23d_chic_n}}
  }
\end{figure} 

\section{\label{sec:23d}The 2D and 3D Systems}

We have computed and analysed series for the specific heat, spin susceptibility
and charge susceptibility  for both the square and simple
cubic lattices. Some representative series are given in Table I.
We choose to present results for the two lattice together so as to highlight
similarities and differences between them.

Figure 3 shows the specific heat at half-filling for various $t/J$ ratios.
The qualitative behaviour is similar to the 1D case, although smaller
values of $t/J$  suffice to produce comparable deviations from the
atomic limit. A two-peak structure is manifest at $t/J=0.4$. Figure 4
shows the dependence of the specific heat on doping (for fixed
$t/J=0.3$). The decrease with decreasing $n$ again confirms that
the high $T$ specific heat is due to conduction electrons.
Figures 5 and 6 show the magnetic susceptibility, at half-filling for
various $t/J$ and at $t/J=0.3$ for various electron densities.
Figure 6 shows a stricking crossover from $n=1$, where $\chi_s$ is
structureless and vanishes as $T\to 0$ to lower densities $n=0.5$ and
0.25 where $\chi_s$ appears to diverge. Evidently $n=0.75$ is near the
critical concentration, as $\chi_s$ appears to start to diverge but then drops to
zero. There is also some apparent structure for the 3D case at $T/J\simeq 0.15$,
which may be an artifact of the analysis.

A mean-field treatment\cite{tsu97} suggests the existence of a finite-temperature
phase transition to a ferromagnetic phase for small $n$, at least for the 3D
case. To explore this we plot the inverse susceptibility versus $T$ in Figure 7.
For both lattices $\chi_s^{-1}$ appears to vanish linearly as $T\to 0$ at
$n=0.5$. For the simple cubic lattice there is some indication that the
curves for $\chi_s^{-1}$ at $n=0.25$ and $n=0.1$ vanish at a small finite
$T$, confirming a finite temperature ferromagnetic phase. 
However the analysis at low temperature is imprecise and does not allow
for an accurate determination of $T_c$.
The data for
the square lattice appear to show no finite temperature
ferromagnetic transition.

Finally in Figures 8 and 9 we show the charge susceptibility versus temperature.
The behaviour is qualitatively similar to the 1D case.

\section{\label{sec4}Conclusions}
We have used thermodynamic perturbation theory to investigate the
antiferromagnetic Kondo lattice model at finite temperatures.
Our calculations focus, in particular, on the specific heat and spin and
charge susceptibilities, and their variation both with the
ratio $t/J$ and with electron concentration. We have presented results for
the linear chain, square lattice, and simple cubic lattice.

Overall, for the parameter region where our series can be analysed with reasonable
precision, the behaviour  of the 3 lattices is qualitatively similar. 
This is to be expected at moderate and high temperatures.
We do,
however, see an indication of a finite temperature ferromagnetic transition in 3-dimensions
for small $n$, consistent with expectations. This
is not seen in 1- and 2-dimensions.

For the linear chain our results are in excellent agreement with the
previous finite-temperature DMRG calculations\cite{shi}, and serve to confirm the
accuracy of both methods. There  are few existing results in higher
dimension and our work should provide a valuable benchmark
for other approaches. While we have not attempted a detailed
comparison on or fit to experiment, this would be possible.
We plan to report results for closed-packed lattices
(triangular and face-centered cubic) and for the ferromagnetic
Kondo lattice model elsewhere.

\appendix
\subsection*{Appendix}

The following high temperature expansions up to order $K^4$ 
have been obtained from the
general results. The series are expressed in terms of $K=\beta J/4$.
The parameters are $\lambda = t/J$, $z=$coordination number,
$p_4$ is weak embedding constant of square cluster, which has value
0, 1, 3, and 12 for 1D, square lattice, simple cubic lattice, and
BCC lattice, respectively, and $z_2= 2 ( 4\,p_4 + z + {z^2})/3$.

Fugacity $\zeta$:
\bea
\zeta &=& 
  {\frac{n}{2 - n}} - {\frac{ n (1-n) }{2 - n}}
       {\Big [} 
        ( {\frac{3} {2}} + 8 z {\lambda}^2 ) K^2 
       + {K^3} \nonumber \\
       &&      -
          {\frac{1} {8}} (11 - 27 n + 9 n^2) K^4
      -  4 z ( 4 - 3 n  ) {{\lambda}^2} K^4 \nonumber \\
&&    -     32 z ( 1 - 6 n + 3 n^2 ) {{\lambda}^4} K^4 - 
               32 z_2 ( 1 +  n  - n^2  )  {{\lambda}^4} K^4 {\Big ] } + O(K^5)
\eea

Internal energy:
\bea
u/J &=& -
  {\frac{1} {32}}  n ( 2 - n ) 
      {\Big [}
         ( 12 + 64 z {{\lambda}^2} ) K 
       + 12 {K^2}
            -  ( 4 - 18 n + 9 {n^2} ) K^3 - 64 z {{\lambda}^2} K^3  \nonumber \\
&& -   256 z ( 2 - 6 n + 3 {n^2} ) {{\lambda}^4} K^3 - 256 z_2 n (
              2  - n)  {{\lambda}^4} K^3 - 5 (4 - 6 n + 3 n^2 ) K^4 - 80 z {{\lambda}^2} K^4
                {\Big ]} + O(K^5)
\eea

Specific heat:
\bea
C_v/k_B &=&
  n \left( 2 - n \right)   {K^2}
   {\Big [} ( {\frac{3}{2}}  + 8 z {{\lambda}^2} ) + 3 K
   - {\frac{3}{8}} ( 4 - 18 n + 9 n^2) K^2       
        - 24 z {{\lambda}^2} K^2  \nonumber \\
&&        - 96 z ( 2 - 6 n + 3 n^2) {{\lambda}^4} K^2
         - 96 z_2 n (2-n )  {{\lambda}^4} K^2 
         {\Big ]} + O(K^5)
\eea

Magnetic susceptibility:
\bea
\beta^{-1} \chi_s &=&
  {\frac{1} {4}}  + {\frac{1} {8}} n ( 2 - n )   
       {\Big [} 1 - 2 K  - {\frac{1} {4}} ( 8 - 6 n + 3 {n^2}) {K^2}  -
            4 z n ( 2 - n )   {{\lambda}^2}  {K^2} \nonumber \\
&&   +  {\frac{1} {3}} ( 2 - 6 n + 3 {n^2})  {K^3}  +
            8  z n ( 2 - n )  {{\lambda}^2} {K^3} \nonumber \\
&&   +  {\frac{1} {24}} (80 - 180 n + 198 {n^2} - 108 {n^3} + 
                27 {n^4} ) {K^4} +
            {\frac{2}{3}} z  ( 12 + 10 n - 41 {n^2} + 36 {n^3} - 9 {n^4} ) {{\lambda}^2} K^4   \nonumber \\                
&&   + 16 z  n ( 2 - n ) ( 7 - 18 n + 9 {n^2} ){{\lambda}^4} K^4
              - 32 z_2 n (2-n) (1 -  4 n  +2  {n^2} )   {{\lambda}^4} K^4
           {\Big ]}  + O(K^5)
%
%
\eea

Charge susceptibility:
%
%
%
%
%
\bea
\beta^{-1} \chi_c &=&
 {\frac{1}{2}}  n ( 2 - n )  - 
    n^2 (2-n)^2 
       {\Big[}   ( {\frac{3}{8}} + 2 z {{\lambda}^2} ) {K^2}  
       + {\frac{1}{4}} {K^3}  \nonumber \\
&&  -   {\frac{1}{16}} ( 10 - 18 n + 9 {n^2}) K^4 
    -  z ( 1 + 6 n - 3 {n^2} ) {{\lambda}^2} K^4  \nonumber \\
&& - 8 z ( 7 - 18 n  + 9 n^2  ) {{\lambda}^4} K^4 
   + 16 z_2 ( 1 - 4 n + 2 n^2) {{\lambda}^4} K^4 {\Big ]}
    + O(K^5)
\eea

\begin{acknowledgments}
This work  forms part of 
a research project supported by a grant
from the Australian Research Council.
The computations were performed on an AlphaServer SC
 computer. We are grateful for the computing resources provided
 by the Australian Partnership for Advanced Computing (APAC)
National Facility.
\end{acknowledgments}

\newpage
\bibliography{basename of .bib file}


\begin{table*}
\squeezetable
\caption{Series coefficients for  the internal energy $u$, specific heat $C_v$, 
magnetic susceptibility $\chi_s$ and charge magnetic susceptibility $\chi_c$ 
at $\beta J=1$ and electron densities $n=1,0.5 $ 
for the linear chain, the
square lattice and the simple cubic lattice. 
Nonzero coefficients $(t/J)^r$
up to order $r=8$  are listed.}\label{tab_ser}
\begin{ruledtabular}
\begin{tabular}{|r|ll|ll|ll|} 
\multicolumn{1}{|c|}{} & \multicolumn{2}{c|}{linear chain} & \multicolumn{2}{c|}{square lattice}
& \multicolumn{2}{c|}{simple cubic lattice} \\
\multicolumn{1}{|c|}{$r$} & \multicolumn{1}{c}{$n=1$} & \multicolumn{1}{c|}{$n=0.5$} 
 & \multicolumn{1}{c}{$n=1$} & \multicolumn{1}{c|}{$n=0.5$} 
  & \multicolumn{1}{c}{$n=1$} & \multicolumn{1}{c|}{$n=0.5$} \\
\hline
\multicolumn{1}{|c|}{}&\multicolumn{6}{c|}{$u/J$}\\
  0 & -1.18727275$\times 10^{-1}$ & -8.78192521$\times 10^{-2}$ & -1.18727275$\times 10^{-1}$ & -8.78192521$\times 10^{-2}$ & -1.18727275$\times 10^{-1}$ & -8.78192521$\times 10^{-2}$ \\
  2 & -9.19083667$\times 10^{-1}$ & -6.90488388$\times 10^{-1}$ & -1.83816733       & -1.38097678       & -2.75725100       & -2.07146516       \\
  4 & ~2.02166868$\times 10^{-1}$ & ~1.94156599$\times 10^{-1}$ & ~1.23945460       & ~8.51133343$\times 10^{-1}$ & ~3.11186319       & ~1.97093023       \\
  6 & -5.81316838$\times 10^{-2}$ & -6.82352222$\times 10^{-2}$ & -1.24540759       & -8.52057857$\times 10^{-1}$ & -5.90066730       & -2.95078662       \\
  8 & ~1.71426415$\times 10^{-2}$ & ~2.25662666$\times 10^{-2}$ & ~1.38876432       & ~9.81227389$\times 10^{-1}$ & ~1.32149047$\times 10^{1}$ & ~5.06962222       \\
\hline
\multicolumn{1}{|c|}{}&\multicolumn{6}{c|}{$C_v/k_B$}\\
  0 & ~1.44045799$\times 10^{-1}$ & ~1.03910138$\times 10^{-1}$ & ~1.44045799$\times 10^{-1}$ & ~1.03910138$\times 10^{-1}$ & ~1.44045799$\times 10^{-1}$ & ~1.03910138$\times 10^{-1}$ \\
  2 & ~7.42355042$\times 10^{-1}$ & ~5.62969188$\times 10^{-1}$ & ~1.48471008       & ~1.12593838       & ~2.22706513       & ~1.68890757       \\
  4 & -5.05380430$\times 10^{-1}$ & -4.99348745$\times 10^{-1}$ & -3.16863938       & -2.18170564       & -7.98977684       & -5.04707069       \\
  6 & ~2.39575311$\times 10^{-1}$ & ~2.94484197$\times 10^{-1}$ & ~5.37297355       & ~3.70460027       & ~2.57521720$\times 10^{1}$ & ~1.27479126$\times 10^{1}$ \\
  8 & -9.61839414$\times 10^{-2}$ & -1.34812280$\times 10^{-1}$ & -8.40726263       & -6.02787617       & -8.15089318$\times 10^{1}$ & -3.07784724$\times 10^{1}$ \\
\hline
\multicolumn{1}{|c|}{}&\multicolumn{6}{c|}{$\beta^{-1}\chi_s$}\\
  0 & ~3.02552920$\times 10^{-1}$ & ~2.88871928$\times 10^{-1}$ & ~3.02552920$\times 10^{-1}$ & ~2.88871928$\times 10^{-1}$ & ~3.02552920$\times 10^{-1}$ & ~2.88871928$\times 10^{-1}$ \\
  2 & -2.69346503$\times 10^{-2}$ & -1.34119052$\times 10^{-2}$ & -5.38693006$\times 10^{-2}$ & -2.68238104$\times 10^{-2}$ & -8.08039510$\times 10^{-2}$ & -4.02357157$\times 10^{-2}$ \\
  4 & ~1.33858987$\times 10^{-2}$ & ~7.41933727$\times 10^{-3}$ & ~8.14693283$\times 10^{-2}$ & ~2.59321663$\times 10^{-2}$ & ~2.04250289$\times 10^{-1}$ & ~5.55384870$\times 10^{-2}$ \\
  6 & -6.15230377$\times 10^{-3}$ & -3.94194796$\times 10^{-3}$ & -1.25846811$\times 10^{-1}$ & -2.89978905$\times 10^{-2}$ & -5.88972872$\times 10^{-1}$ & -5.65349137$\times 10^{-2}$ \\
  8 & ~2.69069927$\times 10^{-3}$ & ~1.92550252$\times 10^{-3}$ & ~1.94026824$\times 10^{-1}$ & ~3.56078818$\times 10^{-2}$ & ~1.78779922$\times 10^{0}$ & -3.46409964$\times 10^{-2}$ \\
\hline
\multicolumn{1}{|c|}{}&\multicolumn{6}{c|}{$ \beta^{-1} \chi_c$}\\
  0 & ~4.73182256$\times 10^{-1}$ & ~3.60327615$\times 10^{-1}$ & ~4.73182256$\times 10^{-1}$ & ~3.60327615$\times 10^{-1}$ & ~4.73182256$\times 10^{-1}$ & ~3.60327615$\times 10^{-1}$ \\
  2 & -2.14752162$\times 10^{-1}$ & -1.24947446$\times 10^{-1}$ & -4.29504323$\times 10^{-1}$ & -2.49894893$\times 10^{-1}$ & -6.44256485$\times 10^{-1}$ & -3.74842339$\times 10^{-1}$ \\
  4 & ~9.57436401$\times 10^{-2}$ & ~6.51283684$\times 10^{-2}$ & ~5.88984198$\times 10^{-1}$ & ~2.45225449$\times 10^{-1}$ & ~1.47972167       & ~5.40291243$\times 10^{-1}$ \\
  6 & -3.93249095$\times 10^{-2}$ & -3.25845479$\times 10^{-2}$ & -8.44234559$\times 10^{-1}$ & -2.81650591$\times 10^{-1}$ & -4.00301487       & -7.06439879$\times 10^{-1}$ \\
  8 & ~1.51554530$\times 10^{-2}$ & ~1.49996060$\times 10^{-2}$ & ~1.22035314       & ~3.56665468$\times 10^{-1}$ & ~1.16033653$\times 10^{1}$ & ~5.01750366$\times 10^{-1}$ \\
\end{tabular}                                                        
\end{ruledtabular}
\end{table*}

\end{document}